\documentclass[twocolumn]{aastex631}

\received{May, 2022}
\revised{--}
\accepted{--}

\submitjournal{ApJ}

\shorttitle{Detailed abundances of stars in the outskirts of Tucana II}
\shortauthors{Chiti et al.}

\newcommand{\teff}{\ensuremath{T_\mathrm{eff}}}
\newcommand{\logg}{\ensuremath{\log\,g}}

\usepackage[thinc]{esdiff}

\begin{document}

\title{Detailed chemical abundances of stars in the outskirts of the Tucana II ultra-faint dwarf galaxy\footnote{This paper includes data gathered with the 6.5 meter Magellan Telescopes located at Las Campanas Observatory, Chile.}}

\correspondingauthor{Anirudh Chiti}
\email{achiti@uchicago.edu}

\author[0000-0002-7155-679X]{Anirudh Chiti}
\affil{Department of Astronomy \& Astrophysics, University of Chicago, 5640 S Ellis Avenue, Chicago, IL 60637, USA}
\affil{Kavli Institute for Cosmological Physics, University of Chicago, Chicago, IL 60637, USA}

\author[0000-0002-2139-7145]{Anna Frebel}
\affiliation{Department of Physics and Kavli Institute for Astrophysics and Space Research, Massachusetts Institute of Technology, Cambridge, MA 02139, USA}

\author[0000-0002-4863-8842]{Alexander P. Ji}
\affil{Department of Astronomy \& Astrophysics, University of Chicago, 5640 S Ellis Avenue, Chicago, IL 60637, USA}
\affil{Kavli Institute for Cosmological Physics, University of Chicago, Chicago, IL 60637, USA}

\author[0000-0001-9178-3992]{Mohammad K.\ Mardini}
\affiliation{Kavli Institute for the Physics and Mathematics of the Universe (WPI), The University of Tokyo Institutes for Advanced Study, and The University of Tokyo, Kashiwa, Chiba 277-8583, Japan}
\affiliation{Institute for AI and Beyond, The University of Tokyo, 7-3-1 Hongo, Bunkyo-ku, Tokyo 113-8655, Japan}
\affiliation{Department of Physics and Kavli Institute for Astrophysics and Space Research, Massachusetts Institute of Technology, Cambridge, MA 02139, USA}
\affiliation{Joint Institute for Nuclear Astrophysics -- Center for the Evolution of the Elements (JINA-CEE), USA}

\author[0000-0002-4669-9967]{Xiaowei Ou}
\affiliation{Department of Physics and Kavli Institute for Astrophysics and Space Research, Massachusetts Institute of Technology, Cambridge, MA 02139, USA}

\author[0000-0002-4733-4994]{Joshua D. Simon}
\affiliation{Observatories of the Carnegie Institution for Science, 813 Santa Barbara St., Pasadena, CA 91101, USA}

\author[0000-0003-4624-9592]{Helmut Jerjen} 
\affiliation{Research School of Astronomy and Astrophysics, Australian National University, Canberra, ACT 2611, Australia}

\author[0000-0002-6658-5908]{Dongwon Kim}
\affiliation{Department of Convergence Medicine, Asan Medical Institute of Convergence Science and Technology, Asan Medical Center, University of Ulsan College of Medicine, Seoul, Republic of Korea.}

\author[0000-0002-7900-5554]{John E. Norris}
\affiliation{Research School of Astronomy and Astrophysics, Australian National University, Canberra, ACT 2611, Australia}

\begin{abstract}

We present chemical abundances and velocities of five stars between 0.3\,kpc to 1.1\,kpc from the center of the Tucana II ultra-faint dwarf galaxy (UFD) from high-resolution Magellan/MIKE spectroscopy. 
We find that every star is deficient in metals ($-3.6<$ [Fe/H] $< -1.9$) and in neutron-capture elements as is characteristic of UFD stars, unambiguously confirming their association with Tucana II. 
Other chemical abundances (e.g., C, iron-peak) largely follow UFD trends and suggest that faint core-collapse supernovae (SNe) dominated the early evolution of Tucana II.
We see a downturn in [$\alpha$/Fe] at [Fe/H] $\approx -2.8$, indicating the onset of Type Ia SN enrichment and somewhat extended chemical evolution.
The most metal-rich star has strikingly low [Sc/Fe] = $-1.29 \pm 0.48$ and [Mn/Fe] = $-1.33 \pm 0.33$, implying significant enrichment by a sub-Chandrasekhar mass Type Ia SN. 
We do not detect a radial velocity gradient in Tucana II ($\text{d}v_{\text{helio}}/\text{d}\theta_1=-2.6^{+3.0}_{-2.9}$\,km\,s$^{-1}$\,kpc$^{-1}$) reflecting a lack of evidence for tidal disruption, and derive a dynamical mass of $M_{1/2}\, (r_h) = 1.6^{+1.1}_{-0.7}\times 10^6$\,M$_{\odot}$. 
We revisit formation scenarios of the extended component of Tucana II in light of its stellar chemical abundances.
We find no evidence that Tucana II had abnormally energetic SNe, suggesting that if SNe drove in-situ stellar halo formation then other UFDs should show similar such features.
Although not a unique explanation, the decline in [$\alpha$/Fe] is consistent with an early galactic merger triggering later star formation.
Future observations may disentangle such formation channels of UFD outskirts. 

\end{abstract}


\section{Introduction} 
\label{sec:intro}

The Milky Way's ultra-faint dwarf satellite galaxies (UFDs; $L \lesssim 10^5 L_{\odot}$, \citealt{bb+17, s+19}) are nearby systems that formed at high redshifts \citep[$z \gtrsim 6$;][]{btg+14, cmr+21, sbd+21}, making them local laboratories for studies of early galaxy formation.
Characteristics of UFDs, including their frequency, size, and mean metallicity, are linked to our understanding of early feedback \citep{apr+19}, early galactic assembly \citep{rpa+19,tyf+21}, reionization and  Milky Way assembly \citep[e.g.,][]{mk+21}, and the nature of dark matter \citep[e.g.,][]{nba+21}. 
UFDs are also of broader astrophysical importance, as they are the most dark matter dominated stellar systems known ($M/L \gtrsim100\,M_{\odot}/L_{\odot}$; \citealt{s+19}) and their most metal-poor stars likely reflect yields from the supernovae of the first stars \citep{jbb+21, rss+21}. 

Detailed chemical abundances of stars in UFDs, in particular, have been impactful in constraining early galactic and chemical evolution. 
For example, the highly r-process enhanced stars in the Reticulum II UFD \citep{jfs+16, rmb+16}, coupled with the dearth of neutron-capture elements in other UFDs \citep[e.g.,][]{jsf+19}, have isolated the site of early r-process nucleosynthesis to be neutron star mergers or a rare class of supernovae \citep{jfc+16, vpr+20}. 
Whether or not a UFD displays a flat $\alpha$-abundance trend with [Fe/H] (as seen in e.g., the Segue 1 UFD with absolute magnitude $M_{\text{V}} = -1.30 \pm 0.73$; \citealt{mcs+18}) diagnoses whether these systems are the result of short episodes of star formation and simple, early chemical enrichment driven by core-collapse supernovae \citep{fb+12, fsk+14, vmv+14, wbs+15, rss+21}.
The existence (or lack) of stars with [Fe/H] $< -4.0$ is related to the prevalence of external chemical enrichment from neighboring minihalos \citep{sst+15, jbb+17}.
And recently, differences in the [Mg/Ca] vs. [Fe/H] trends between UFDs associated with the Large Magellanic Cloud (LMC) and those associated with the Milky Way have hinted at environmental variations in the early evolution of galaxies \citep{jls+20}.
However, only a handful of stars (generally $\lesssim$ 5) per UFD are typically observable with high-resolution spectroscopy,
making a detailed chemical abundance analysis of more UFD stars crucial for a comprehensive understanding of their evolution. 

Tucana II is a particularly interesting UFD for such studies, due to the recent detection of an extended component of member stars out to 1.1\,kpc from its center by \citet{cfs+21}. 
No evidence was found that the extended nature is due to tidal disruption, suggesting that these stars instead trace the underlying extended dark matter halo of the UFD. 
The detailed chemical characterization of stars in this spatial regime is particularly intriguing, not only to increase the overall sample of UFD stars with abundance information but also to shed light on the formation of such an extended structure around a UFD.
Indeed, recent simulation work has suggested that this feature around Tucana II (and any such features around UFDs in general) can be linked to assembly events between building block galaxies \citep{rpa+19, tyf+21}, although in-situ formation, e.g., through early feedback, is also a possibility.

Differentiating the formation channels of the extended, outer region of Tucana II is principally possible with results from high-resolution spectroscopy. 
For example, a marked distinction between the chemical abundances of the inner and outer stars, or inconsistencies when modeling the chemical evolution of the full sample, can indicate an ex-situ origin of the extended stars. 
Moreover, the higher velocity precision afforded by high-resolution spectroscopy can better constrain whether the extended component is bound, or formed through tidal disruption, by means of tighter constraints on whether a radial velocity gradient exists \citep[e.g.,][]{lsk+18}. 

In this paper, we present results from high-resolution Magellan/MIKE spectroscopy of the five stars in the outskirts (0.3\,kpc to 1.1\,kpc from the center) of Tucana II, providing a suite of chemical abundances (e.g., C, $\alpha$-elements, neutron-capture elements) and a $\sim3$x improvement in the radial velocity precision relative to our previous medium-resolution spectroscopic analyses presented in \citet{cfs+21}.
In Section~\ref{sec:obs}, we describe our new observations; in Section~\ref{sec:analysis}, we outline our radial velocity and chemical abundance analyses; in Section~\ref{sec:results}, we present the detailed chemical abundance signatures of these stars; in Section~\ref{sec:discussion}, we comment on the dynamical state and early evolution of Tucana II; and in Section~\ref{sec:conclusion}, we conclude.

\section{Sample Selection \& Observations}
\label{sec:obs}

\begin{deluxetable*}{lllllllll} 
\tablecolumns{8}
\tablecaption{\label{tab:obs} Observations}
\tablehead{   
  \colhead{Name$\tablenotemark{a}$} &
  \colhead{RA (h:m:s)} & 
  \colhead{DEC (d:m:s)} &
  \colhead{UT Observation Dates} &
  \colhead{Slit size} &
  \colhead{$g$} &
  \colhead{$t_{\text{exp}}$} &
  \colhead{S/N$\tablenotemark{b}$} & 
  \colhead{$v_{\text{helio}}$} \\
  &
  (J2000) &
  (J2000) & 
  &
  &
  (mag) &
  (min) &
  &
  (km/s)   
}
\startdata
TucII-301 & 22:50:45.097 & $-58$:56:20.483 & 2021 Jul 30, Oct 05 & 1\farcs0 & 18.87 & 355 & 12, 22 & $-125.1 \pm 0.9 $\\
TucII-303 & 22:53:05.194 & $-57$:54:27.032 & 2020 Oct 10, Dec 01 & 1\farcs0 & 18.44 & 310 & 15, 33 & $-128.9 \pm 0.9 $\\
TucII-305 & 22:57:46.859 & $-57$:43:39.299 & 2020 Oct 08, 09, 10 & 1\farcs0 & 18.47 & 365 & 19, 40 & $-125.3 \pm 0.9 $\\
TucII-306 & 22:51:37.019 & $-58$:53:37.579 & 2021 Jul 30, Oct 05& 1\farcs0 & 18.38 & 150 & 17, 28 & $-119.1 \pm 0.9 $\\
TucII-309 & 22:49:24.690 & $-58$:20:47.429 & 2020 Oct 10; 2021 Jun 07, Oct 05 & 1\farcs0 & 18.73 & 203 & 15, 28 & $-125.3 \pm 0.9$\\
\enddata
\tablenotetext{a}{Names are as indicated in \citet{cfs+21}}
\tablenotetext{b}{S/N per pixel is listed for 4500\,\AA\,\,and 6500\,\AA}
\end{deluxetable*}

We used Magellan/MIKE \citep{bsg+03} to obtain high-resolution ($R\sim25,000$) spectra of five Tucana II member stars presented in \citet{cfs+21}: TucII-301, TucII-303, TucII-305, TucII-306, and TucII-309. 
These stars lie far from the center of the UFD (from 0.3\,kpc to 1.1\,kpc) and were initially identified as candidate Tucana II members in \citet{cfj+20} due to their low photometric metallicities and surface gravities, as derived from deep imaging of the UFD using the SkyMapper Telescope \citep{ksb+07, bbs+11}.
These stars also displayed \textit{Gaia} DR2 proper motions \citet{gaia+16, gaia+18} that were consistent with the systemic proper motion of Tucana II \citep[e.g.,][]{simon+18, mv+20}, increasing confidence in their association with the galaxy.
Ultimately, a radial velocity and spectroscopic metallicity analysis of these stars in \citet{cfs+21} confirmed them to be members of Tucana II.

The MIKE spectra in this study were obtained using the 1\farcs0 slit and 2x2 binning, which provided wavelength coverage between 3500\,{\AA} and 9000\,{\AA} with a resolution of $R\sim28,000$ on the blue echelle orders and $R\sim22,000$ on the red echelle orders. 
The observations took place in October and December 2020; and in June, July, and October 2021. 
The weather was clear and the seeing was good ($< 1\farcs0)$ on every night of observation except for June 07 2021, during which the telescope closed early due to high wind after 40\,mins of data collection.
The data were reduced using the MIKE \texttt{CarPy} reduction pipeline\footnote{https://code.obs.carnegiescience.edu/mike} \citep{k+03}, following standard reduction procedures.
Full details of the observations, including dates, total exposure times, and signal-to-noise (S/N) values are presented in Table~\ref{tab:obs}.

\begin{figure*}[!htbp]
\centering
\includegraphics[width =\textwidth]{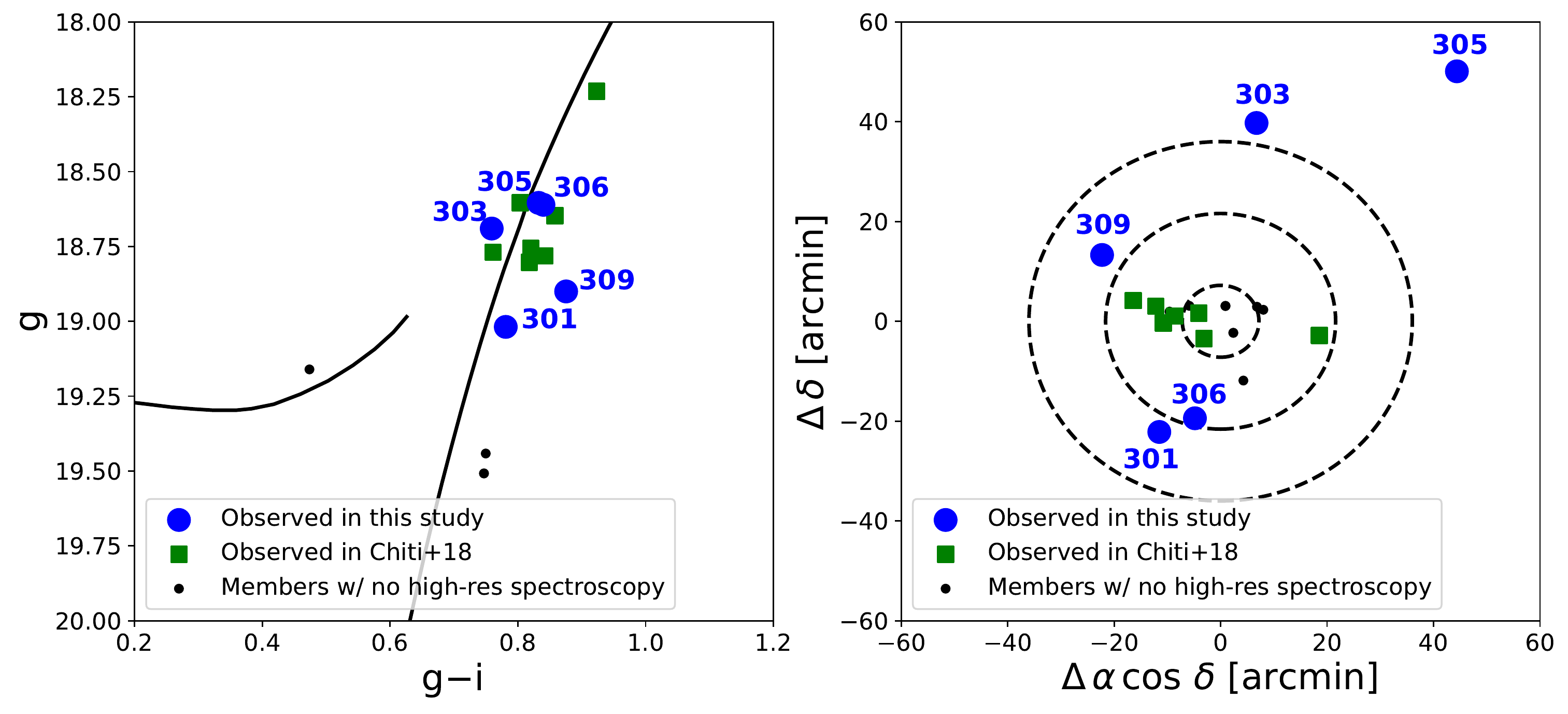}
\caption{Left: Color-magnitude diagram of stars in the Tucana II UFD. Blue circles correspond to stars observed with high-resolution spectroscopy presented in this work; green squares correspond to stars with high-resolution spectroscopy presented in previous works \citep{jfe+16, cfj+18}; black dots correspond to stars with no high-resolution spectroscopy \citep{wmo+16, cfs+21}. A MIST isochrone of 10\,Gyr and [Fe/H] = $-2.2$ \citep{d+16, cdc+16, pbd+11, pcb+13, pms+15, psb+18} is overplotted at the distance modulus of Tucana II \citep[$m-M=18.8$;][]{bdb+15}, for reference. 
Each star observed in this study is labeled by its identifier in Table~\ref{tab:obs}.
Right: Spatial location of stars with respect to the center of Tucana II. 
Dashed circles correspond to 1, 3, and 5 times the half-light radius as presented in \citet{bdb+15}. 
}
\label{fig:cmd}
\end{figure*}

\begin{figure*}[!htbp]
\centering
\includegraphics[width =\textwidth]{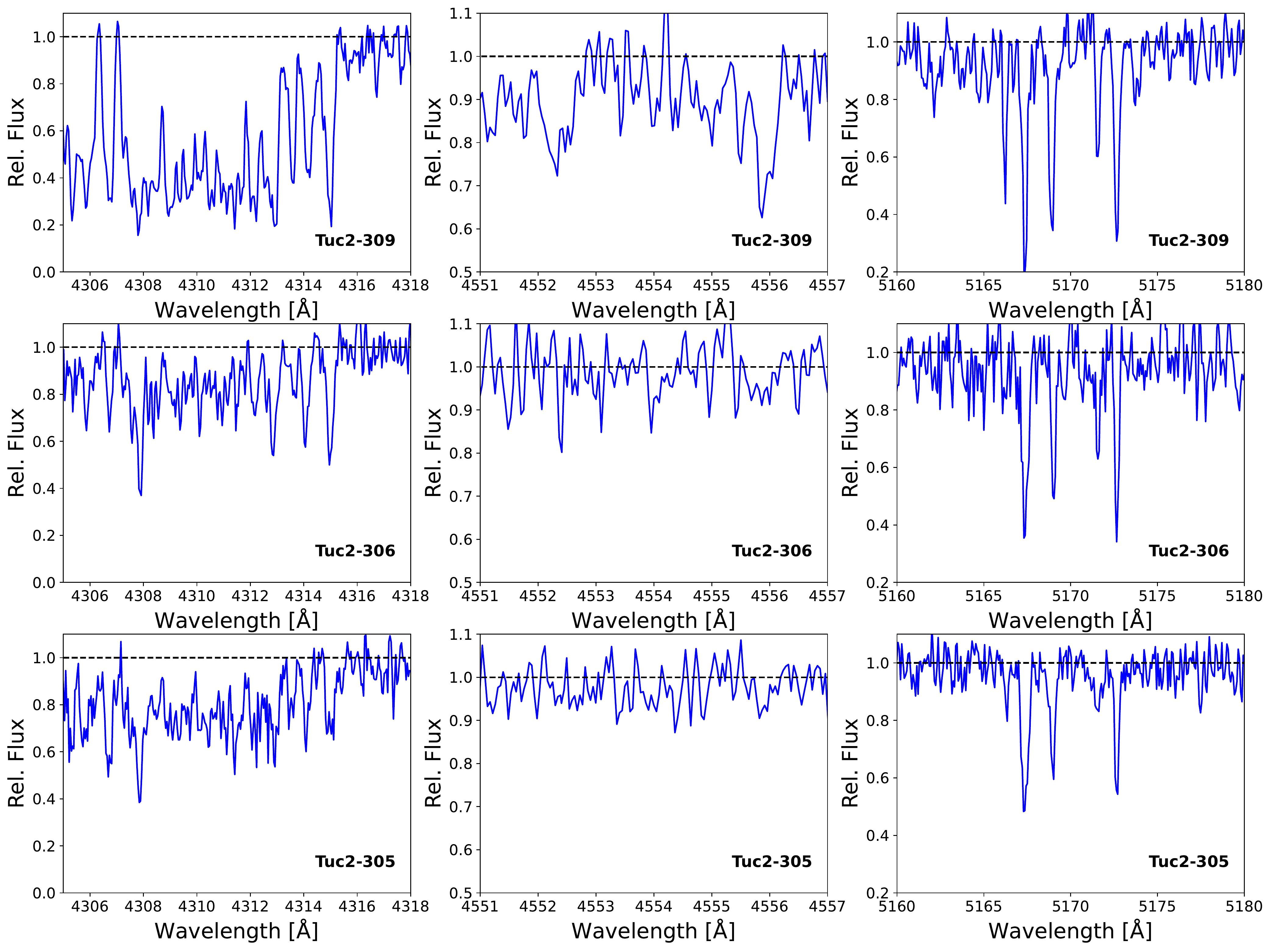}
\caption{Sample regions of our Magellan/MIKE spectra covering the CH absorption band at $\sim4311$\,{\AA} (left panels), the Ba line region at 4554\,{\AA} (center panels), and the Mg\,b region around $5170$\,{\AA} (right panels). 
Tuc2-309, the most metal-rich ([Fe/H] = $-1.93$) star, is plotted in the top panels; TucII-306, with a metallicity of [Fe/H] = $-3.26$, is plotted in the middle panels; and TucII-305, the most metal-poor star ([Fe/H] = $-3.59$), is plotted in the bottom panels.
Note the lack of a detectable Ba feature in these stars, supporting their classification as UFD members (see Section~\ref{sec:ncap}).}
\label{fig:specs}
\end{figure*}

It is worth reiterating the uniqueness of this Tucana II sample which presents the first high-resolution chemical abundance analysis of stars in the outskirts of any UFD that is not known to be tidally disrupting.
This is illustrated in the right panel of Figure~\ref{fig:cmd}; the previously analyzed seven Tucana II member stars \citep{jfe+16, cfj+18} were largely contained within two half-light radii of the galaxy \citep{bdb+15, kbt+15}. 
A color-magnitude diagram of the full Tucana II sample is shown in 
Figure~\ref{fig:cmd} along with its spatial distribution. As can be seen, our new sample is further extended than previously observed stars. As such, we bring the total number of stars with available, detailed chemical abundances in Tucana II to twelve stars, placing it second to only Bootes I \citep{fek+09, nyg+10, llb+11, gnm+13, iaa+14, fng+16} among the entire UFD population. 
Sample regions of our reduced spectra are shown in Figure~\ref{fig:specs}.

\section{Radial Velocity and Chemical Abundance Analysis}
\label{sec:analysis}

\subsection{Derivation of radial velocities}
\label{sec:velocities}

We measured radial velocities following the analysis in \citet{jlh+20}, with a few modifications. 
First, we reduced a separate MIKE spectrum for each night of observation for each star (see Table~\ref{tab:obs} for the list of nights). 
Then, we derived radial velocities independently from each echelle order by cross-correlating the observed data with a high S/N reference spectrum of the very metal-poor red giant HD122563. 
We performed this analysis using the \texttt{measure\_mike\_velocities} function in the \texttt{alexmods}\footnote{https://github.com/alexji/alexmods} package, which we modified to allow a visual inspection of the resulting cross-correlation functions. 
The cross-correlation functions without a clearly separated velocity peak or bimodal velocity peaks were excluded from further analysis.
This step removed orders with e.g., low signal-to-noise or bad continuum normalization.
We then further excluded orders with velocities that were $>3$ sigma outliers relative to the velocities obtained for all orders of a given spectrum.
Finally, we took an inverse-variance weighted mean of the remaining individual-order-velocities (typically from $\sim$10 to $\sim30$ remaining echelle orders, depending on the S/N of the spectrum) to derive a final radial velocity for each MIKE spectrum.
We adopted the weighted standard deviation of the order-velocities as the random uncertainties, which were on the order of $\lesssim0.4\,$km\,s$^{-1}$.
We combined velocities across multiple nights by taking the weighted average, with weights equal to the inverse of the variances.


We also estimated the systematic radial velocity uncertainty using velocities derived from repeat observations of the same star.
Such a sample was readily available as we derived an independent radial velocity from each night of MIKE data.
We find that a systematic uncertainty of 0.9\,km\,s$^{-1}$ needed to be added in quadrature to the random uncertainties for consistency between repeat observations, following the method in \citet{sg+07}. 
The systematic uncertainty dominates the random uncertainties ($< 0.4$\,km\,s$^{-1}$) and drives the overall uncertainty in the velocity measurements.
Final velocities and uncertainties are shown in Table~\ref{tab:obs}, which we derived as the weighted average of the set of nightly radial velocities.
We find no evidence for binarity within the MIKE velocity measurements of our sample of stars.

We compared our MIKE velocities to MagE velocities derived in \citet{cfs+21} to gauge accuracy and further test for binarity.
All of our velocities are in 1$\sigma$ agreement after excluding one star, TucII-309, for which we measure a velocity higher by 8.4\,km\,s$^{-1}$ ($>$ 2$\sigma$ tension). 
This discrepancy suggests that TucII-309 may be in a binary system.
We find an overall systematic offset relative to the MagE velocities of 1.0\,km\,s$^{-1}$ when excluding TucII-309, and an offset of 2.6\,km\,s$^{-1}$ when including it. 

We note that for the dynamical analysis in this paper (Section~\ref{sec:discussion}), we re-derived radial velocities from the MIKE spectra of the 7 additional Tucana II members in \citet{cfj+18} exactly following the procedure described in the first two paragraphs of this Section.
Note that we derive that these velocities have a higher systematic uncertainty of 1.2\,km\,s$^{-1}$, likely due to these data being obtained when MIKE did not have an atmospheric dispersion corrector.
We derive the following radial velocities: $-124.2 \pm 1.2$\,km\,s$^{-1}$ for TucII-006; $-124.4 \pm 1.2$\,km\,s$^{-1}$ for TucII-011; $-126.4 \pm 1.2$\,km\,s$^{-1}$ for TucII-033; $-121.1 \pm 1.2$\,km\,s$^{-1}$ for TucII-052; $-124.5 \pm 1.2$\,km\,s$^{-1}$ for TucII-078; $-126.3 \pm 1.2$\,km\,s$^{-1}$ for TucII-203; $-122.8 \pm 1.2$\,km\,s$^{-1}$ for TucII-206.
The first five of these stars have velocities presented in \citet{wmo+16} from M2FS spectra.
Upon comparing our velocities to those, we find clear evidence for binarity in TucII-078 due to a $> 8$\,km\,s$^{-1}$ difference in the radial velocities. 
Excluding this binary candidate, we find a small but statistically significant offset of $2.5 \pm 0.7$\,km\,s$^{-1}$ between the velocities derived from MIKE and those derived from M2FS in \citet{wmo+16}.
After accounting for this offset, no other stars show evidence for binarity from a comparison between M2FS and MIKE velocities. 

For completeness, we note that two stars observed with MIKE (TucII-006 and TucII-011) have velocities measured from the IMACS instrument in \citet{cfs+21}. 
We find an offset of $2.2$\,km\,s$^{-1}$ between the MIKE velocities of these stars and their IMACS velocities, but the small sample size precludes a robust constraint of this offset.

We include all confirmed members of Tucana II that do not show evidence for binarity in our dynamical analysis in Section~\ref{sec:dynamics}.
We aim to minimize systematics from combining velocities from different instruments through prioritizing measurements from (1) the same spectrograph, (2) spectrographs with the smallest/least evidence for relative velocity offsets, and (3) spectrographs with similar systematic uncertainties (e.g., MIKE and IMACS). 
Accordingly, we include MIKE-based velocities for Tucana II members TucII-006, TucII-011, TucII-033, TucII-052, TucII-203, TucII-206, and all stars in Table~\ref{tab:obs} except TucII-309. 
We use IMACS-based velocities for members TucII-022/Star32, Star12, and Star68 from \citet{cfs+21}. 
We use the M2FS-based velocities of TucII-074 and TucII-085 from \citet{wmo+16}. 
And we use MagE-based velocities for members TucII-310 and TucII-320 from \citet{cfs+21}. 
Note that the binary candidates TucII-309 and TucII-078 are excluded from this analysis. 
The sample for the dynamical analysis in Section~\ref{sec:dynamics} is thus 16 stars. 
A compilation of velocities of confirmed Tucana II members, both from the literature and from this study, is provided in Appendix~\ref{app:vels}.
We note that a velocity offset of +2.5\,km\,s$^{-1}$ is added to the M2FS velocities presented in Appendix~\ref{app:vels} to bring those velocities in agreement with the MIKE velocities.

\subsection{Derivation of stellar parameters and chemical abundances}
\label{sec:abundances}

Stellar parameters and chemical abundances of our stars were derived following standard stellar spectrum analysis techniques \cite[e.g.,][]{fcj+13}.
In particular, we exactly followed the methods used in the previously published MIKE analysis of Tucana II stars \citep[e.g.,][]{cfj+18} to ensure consistency across the entire Tucana II sample. 
We outline the methodology here, for completeness. 

We performed our analysis largely within the Spectroscopy Made Hard (SMH) software package \citep{c+14}, which is a user-friendly Python interface that wraps around e.g., radiative transfer codes and model atmospheres, that are used for stellar spectrum analyses. 
We normalized our spectra using \texttt{run\_continuum.py} in the aforementioned \texttt{alexmods} package (see Section~\ref{sec:velocities}), which fits cubic splines to each spectral order after masking regions of prominent absorption.
Then, individual absorption lines from the line list presented in \citet{rpt+14} were identified in our data.
Equivalent widths for these lines were derived by fitting Gaussians to absorption features.
We used the 2017 version of the MOOG radiative transfer code\footnote{https://github.com/alexji/moog17scat} \citep{s+73} with an updated treatment of scattering \citep{sks+11}, and 1D, $\alpha$-enhanced plane-parallel ATLAS9 model atmospheres \citep{ck+04} to derive our chemical abundances. 

We initially approximated the stellar parameters ($T_{\text{eff}}$, $v_{\text{micro}}$, $\log\,g$) of stars through their relative location on the MIST isochrone in Figure~\ref{fig:cmd} to have a starting point for our iterative analysis. 
Then, these estimates were adjusted until (1) the Fe I abundances showed no trends as a function of reduced equivalent width and excitation potential, and (2) the Fe I and Fe II abundances were in agreement. 
A subsequent temperature correction was applied to $T_{\text{eff}}$ following \citet{fcj+13}, and $v_{\text{micro}}$ and $\log\,g$ were adjusted again until the Fe I abundances showed no trend with reduced equivalent width, and the Fe I and Fe II abundances agreed.
Random uncertainties on the stellar parameters were derived by varying each parameter until the $1\sigma$ scatter in the aforementioned Fe I trends were encompassed.
The total uncertainty on the stellar parameters was derived by adding these random uncertainties in quadrature to the systematic uncertainties, which, as in \citet{cfj+18}, were taken to be 150\,K for $T_{\text{eff}}$, 0.2\,km\,s$^{-1}$ for $v_{\text{micro}}$, and 0.3\,dex for $\log\,g$. 

We derived chemical abundances from molecular line features and dense regions of absorption through spectral synthesis, in which synthetic spectra were generated and chemical abundances were varied until the synthetic spectra matched the observed spectra. 
The line list used for this analysis included the original \citet{rpt+14} list that was used for the equivalent width measurements, in addition to lines from \citet{jfs+16} which incorporated data from \citet{hpc+02}, \citet{dls+03}, \citet{iss+06}, \citet{lds+06}, \citet{lsc+09}, \citet{slc+09}, and \citet{mpv+14}.

The random uncertainties in the chemical abundances were derived exactly following \citet{cfj+18}.
For elements with chemical abundance determinations from $\geq$ 10 absorption features, the random uncertainty was assumed to be the standard deviation of the abundance values. 
For elements with abundances derived from two to ten features, the random uncertainty was derived by multiplying the abundance range by the $k$-statistic \citep{k+62}; and for elements with only one detected absorption feature, the random uncertainty was derived by varying the continuum placement.
If the random uncertainty in an elemental abundance was below that of the uncertainty in the iron abundance, we adopted the uncertainty in the iron abundance to exclude artificially low random uncertainties.
Systematic uncertainties in the chemical abundances were derived by varying each stellar parameter ($T_{\text{eff}}$, $v_{\text{micro}}$, $\log\,g$) by their uncertainty in Table~\ref{tab:stellarparameters}, re-deriving the chemical abundances, and taking the difference between the re-derived chemical abundance and the original abundance as the systematic uncertainty. These systematic uncertainties were added in quadrature to the random uncertainties to derive the total uncertainties.

Our derived chemical abundance values are listed in Table~\ref{tab:abundancetable}, the full uncertainties on these abundances are listed in Table~\ref{tab:uncertainties}, and the individual line measurements are listed in Table~\ref{tab:linetable}. 
We note if certain features were detected, but appear distorted due to e.g., low S/N or were less sensitive to abundance variations, we list abundances derived from those features with a colon in Table~\ref{tab:abundancetable} to denote a highly uncertain value. 
We list all abundances relative to solar abundances from \citet{ags+09}.

\begin{deluxetable}{lllll} 
\tablecolumns{5}
\tablewidth{\columnwidth}
\tablecaption{\label{tab:stellarparameters} Stellar parameters}
\tablehead{   
  \colhead{Name} &
  \colhead{$T_{\text{eff}}$} & 
  \colhead{Log\,$g$} &
  \colhead{v$_{\text{micro}}$} &
  \colhead{[Fe/H]} \\
  \colhead{} &
  \colhead{(K)} & 
  \colhead{(dex)} &
  \colhead{(km s$^{-1}$)} & 
  \colhead{(dex)}
}
\startdata
TucII-301 & $4864\pm241$ & $1.60\pm0.42$ & $1.73\pm0.27$ & $-3.41\pm0.23$ \\
TucII-303 & $5071\pm183$ & $1.55\pm0.81$ & $2.24\pm0.23$ & $-2.74\pm0.17$ \\
TucII-305 & $4810\pm226$ & $1.60\pm0.31$ & $2.30\pm0.26$ & $-3.59\pm0.26$ \\
TucII-306 & $4855\pm215$ & $1.55\pm0.46$ & $2.04\pm0.23$ & $-3.26\pm0.22$ \\
TucII-309 & $4900\pm160$ & $1.75\pm0.76$ & $2.20\pm0.22$ & $-1.93\pm0.22$ \\
\enddata
\end{deluxetable}

\section{The Chemical Abundances of Stars in the Outskirts of Tucana II}
\label{sec:results}

\begin{figure*}[!htbp]
\centering
\includegraphics[width =\textwidth]{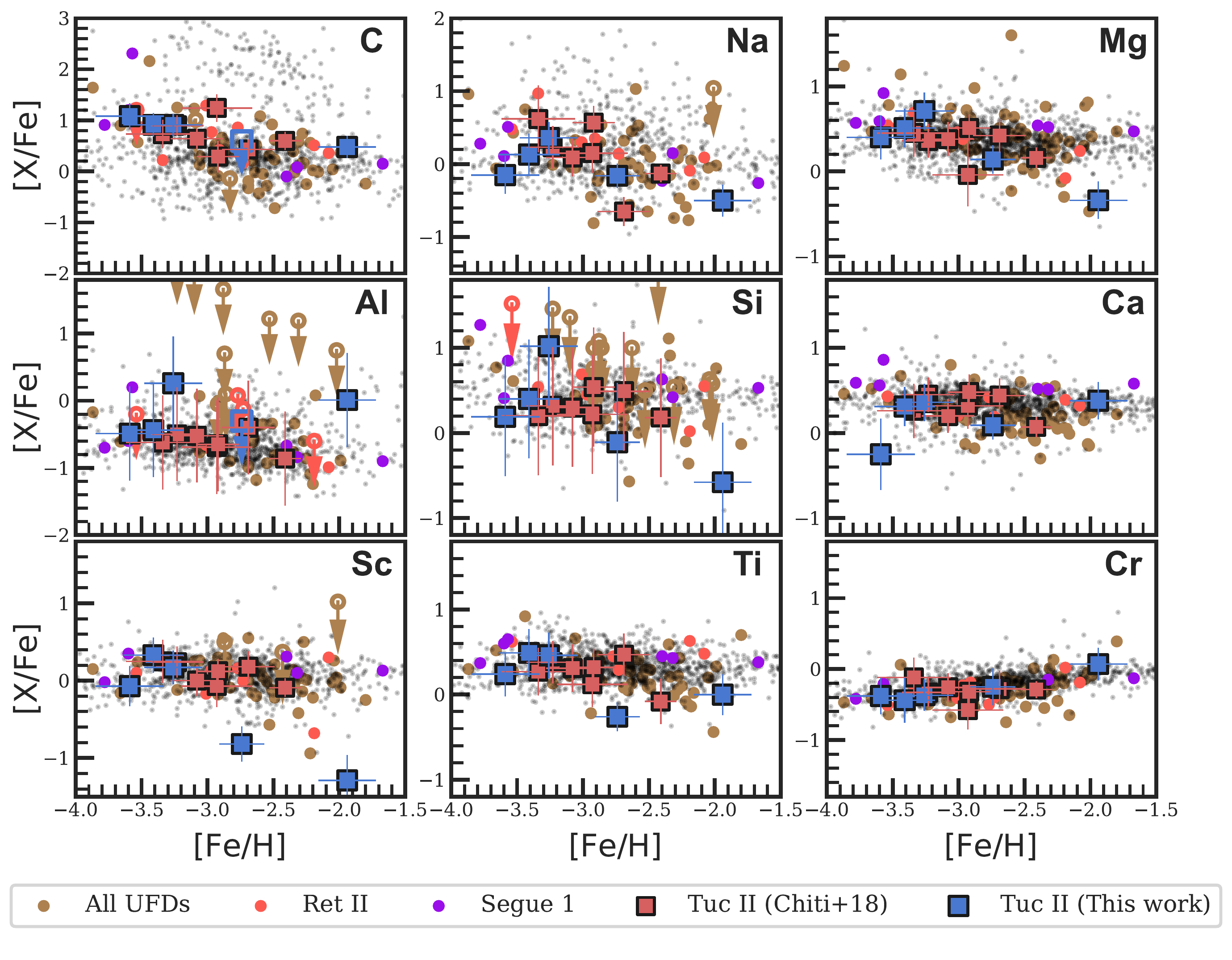}
\caption{[X/Fe] vs. [Fe/H] trends for stars in our Tucana II sample (large blue squares), stars in the inner region of Tucana II (red squares), and other UFDs (brown circles) for carbon, odd-Z elements (Na, Al, Sc), alpha-elements (Mg, Si, Ca, Ti), and Cr. 
The Ret II and Segue 1 UFDs are shown as red circles and orange circles, respectively, due to their interesting chemical signatures (e.g., exceptional r-process enhancement in Ret II, and a flat $\alpha$-element abundance trend in Segue 1). 
The grey data points correspond to abundances for Milky Way (MW) halo stars from the compilation of \citet{af+18}. 
The abundances largely follow the overall UFD trends, with the exception of a low Ca abundance for the most metal-poor star in the system, and a low Mg and Sc abundance in the most metal-rich star in the system.
UFD chemical abundance data is from \citet{kmg+08, fek+09, nwg+10, nyg+10, sfm+10, fsg+10, llb+11, gnm+13, kfa+13, fsk+14, iaa+14, rk+14, fmb+16, jfs+16, fng+16, jfs+16, rmb+16, hsm+17, kcs+17, cfj+18, nml+18, ssf+18, jsf+19, mhs+19, hms+20, jls+20}.}
\label{fig:grid1}
\end{figure*}

\begin{figure*}[!htbp]
\centering
\includegraphics[width =\textwidth]{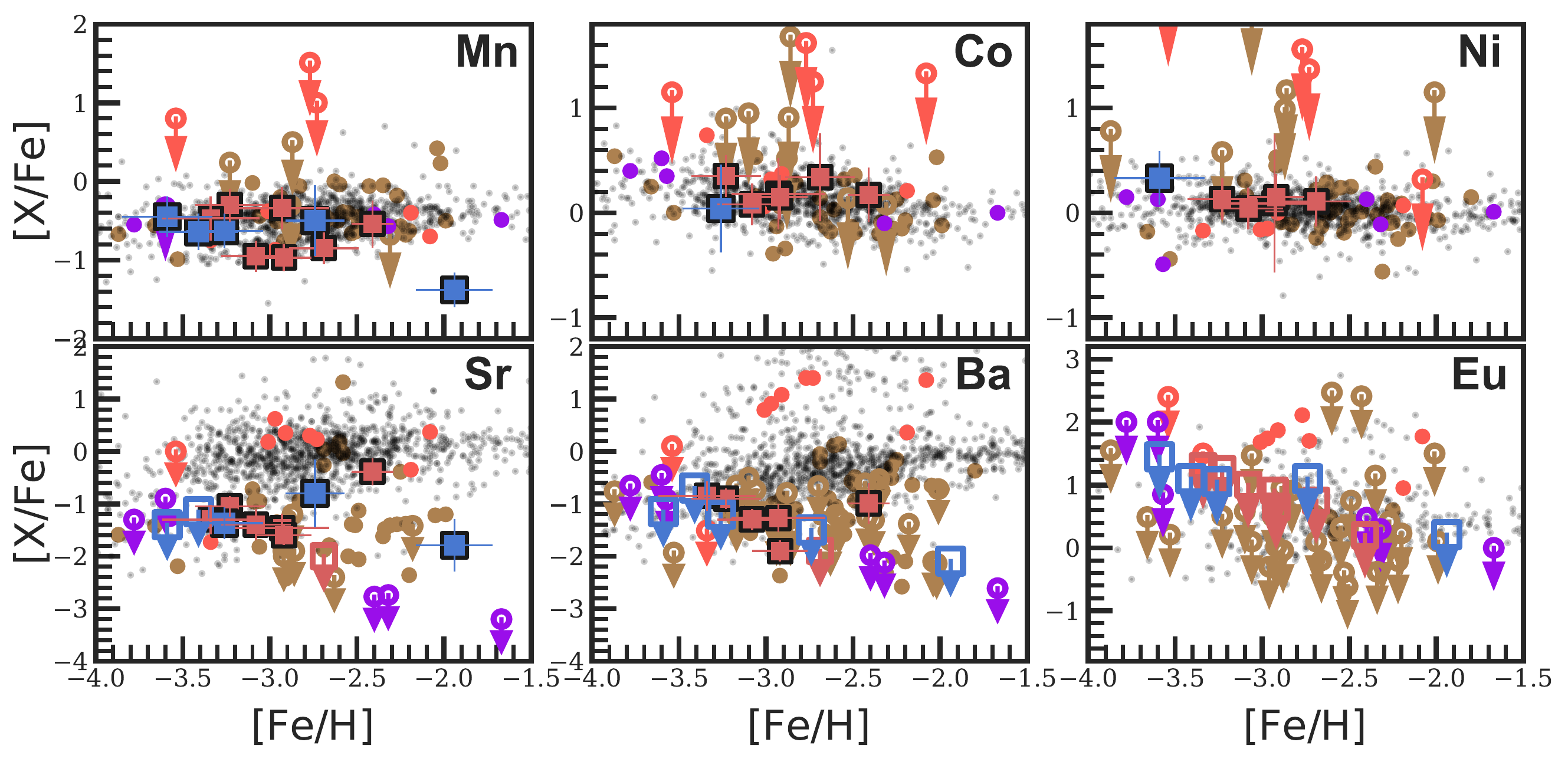}
\caption{[X/Fe] vs. [Fe/H] trends of stars in our Tucana II sample (large blue circles), stars in the inner region of Tucana II (red circles), and other UFDs for the iron-peak elements Mn, Co, and Ni, and the neutron-capture elements Sr, Ba, and Eu. 
The UFD data are plotted with the same legend as in Figure~\ref{fig:grid1}. 
Our newly characterized Tucana II stars show deficiency in Sr and Ba abundances, as is typical for UFD stars, confirming their association with the system. 
The iron-peak abundances largely follow trends seen in other UFDs and the MW halo, except for the most metal-rich Tucana II star that appears to be deficient in Mn. }
\label{fig:grid2}
\label{12}
\end{figure*}

In this section, we provide an overview of the chemical abundances of the five Tucana II stars in our sample.
Specifically, we note the most prominent absorption features from which these abundances were derived, indicate any outlying chemical abundances, and present our results in the context of other UFD stars and Milky Way (MW) halo stars. 
The discussion in this section is purely descriptive, and we refer the reader to Section~\ref{sec:discussion} for interpretations and a discussion of the implications of the abundance trends on the early evolution of Tucana II.
The full suite of chemical abundances of our Tucana II stars are plotted in [X/Fe] vs. [Fe/H] space as large blue squares in Figures~\ref{fig:grid1} and~\ref{fig:grid2}, along with the abundances of other stars in Tucana II (red squares; \citealt{cfj+18}), UFD stars (colored points; see references in Figure~\ref{fig:grid1}) and the MW halo (gray points). 

\begin{figure}[!htbp]
\centering
\includegraphics[width =\columnwidth]{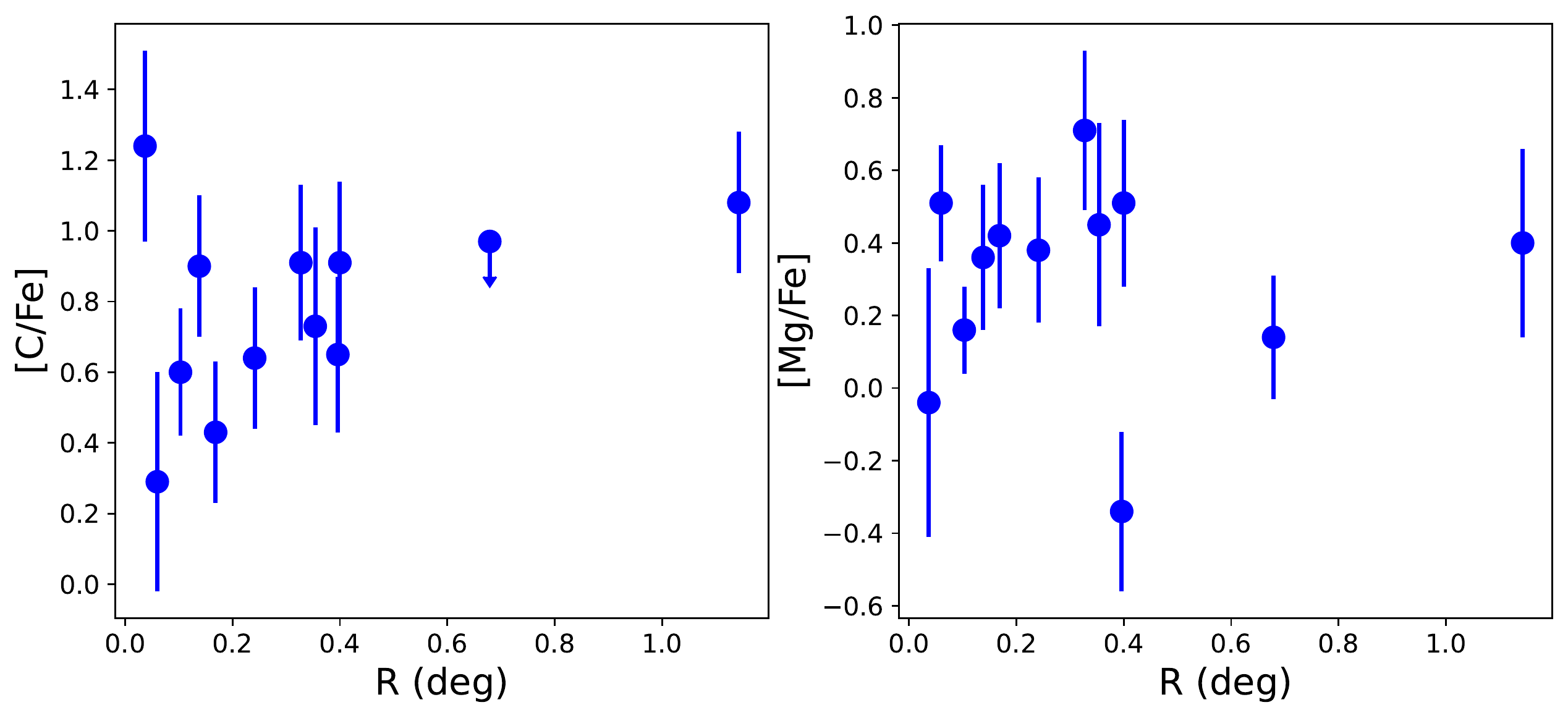}
\caption{Left: [C/Fe] of Tucana II stars as a function of their distance from the center of the UFD. 
Right: [Mg/Fe] of Tucana II stars as a function of their distance from the center of the UFD. 
The spatial distribution of both of these abundances do not show strong statistical evidence for spatial correlations (see Sections~\ref{sec:carbon} and~\ref{sec:alpha}).}
\label{fig:spatialabundances}
\label{12}
\end{figure}

\begin{figure}[!htbp]
\centering
\includegraphics[width =\columnwidth]{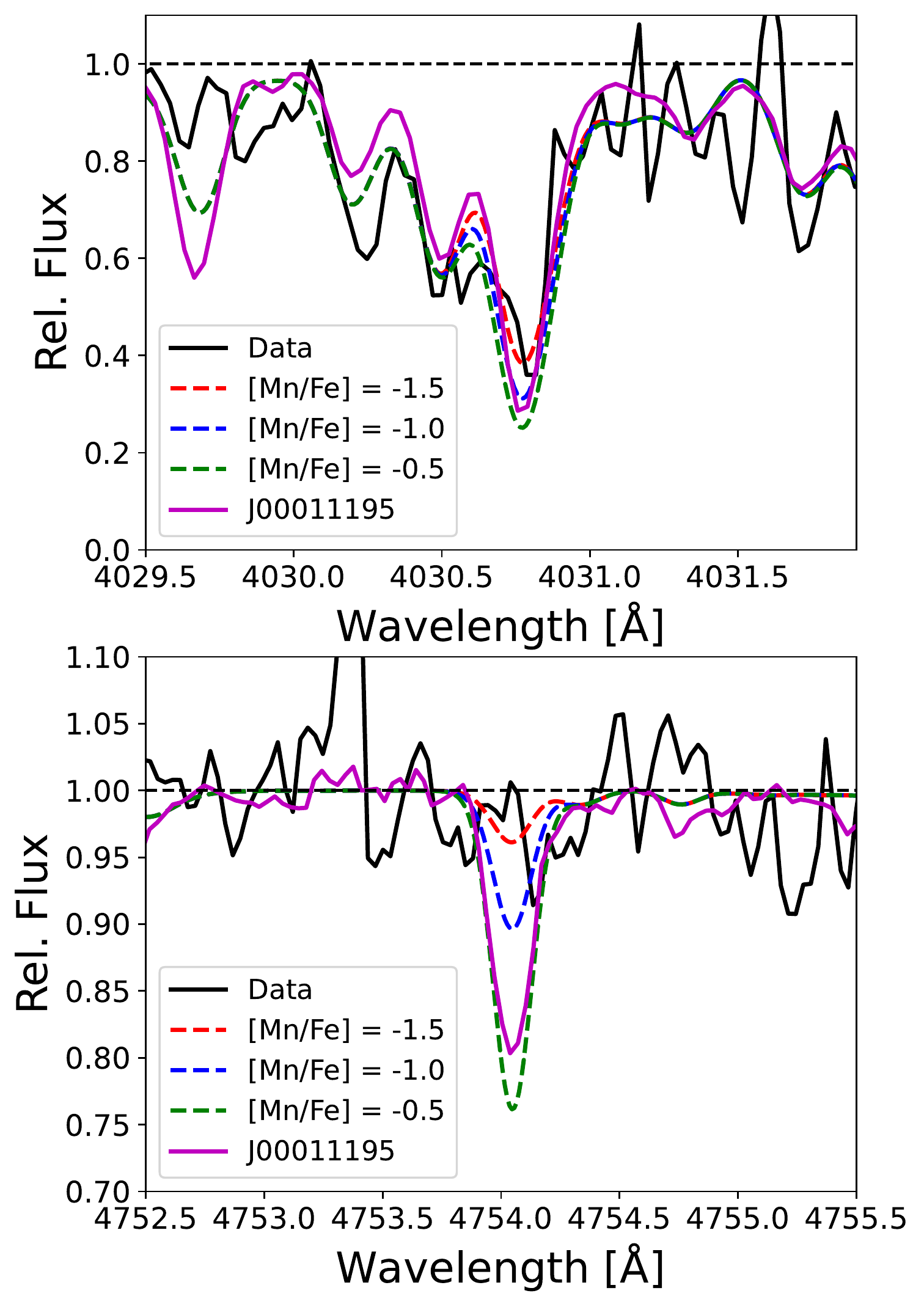}
\caption{Top panel: Spectrum of TucII-309 (black line) around the Mn 4030.7\,{\AA} absorption feature, with synthetic spectra at [Mn/Fe] = $-1.5, -1.0$ and $-0.5$ overlaid along with a reference MIKE spectrum of the metal-poor halo star 2MASS\,J00011195+0321051 (purple) from \citet{erf+20}.
2MASS\,J00011195+0321051 has similar stellar parameters (\teff = 4909\,K, $\log\,g$ = 1.75, $v_{\text{micro}} = 1.85$, [Fe/H] = $-1.99$; \citealt{erf+20}) to TucII-309 but a [Mn/Fe] = $-0.36$, in line with the halo trend (see Figure~\ref{fig:grid2}).
Bottom panel: Same as top panel, but around the Mn 4754\,{\AA} absorption feature. The 2MASS\,J00011195+0321051 Mn 4754\,{\AA} absorption feature is clearly visible, but the feature is not detected in TucII-309, demonstrating the low Mn abundance of TucII-309.
Note that the Mn abundance from the 4754\,{\AA} feature in \citet{erf+20} for 2MASS\,J00011195+0321051 is [Mg/H] = $-2.51$, consistent with the above synthesis.}
\label{fig:MnSc}
\label{12}
\end{figure}

\subsection{Carbon}
\label{sec:carbon}

The carbon abundances of our stars were derived using spectral synthesis over the CH absorption bands at 4313\,{\AA} and 4323\,\AA.
We then corrected these carbon abundances for the evolutionary state of our stars, following \citet{pfb+14}, to account for the carbon depletion in stars as they begin ascending the red giant branch.
Both the uncorrected and corrected carbon abundances are listed in the first two rows of Table~\ref{tab:abundancetable}.
We clearly detect carbon absorption features in the spectrum of each star, except for TucII-303, for which we place a 2$\sigma$ upper limit on the carbon abundance. 

The carbon abundances of Tucana II stars largely track what is seen in other UFDs and the MW halo (see the top left panel in Figure~\ref{fig:grid1}).
Three of the stars in our sample (including the two most metal-poor Tuc II stars) can be considered carbon-enhanced metal-poor stars (CEMP; [C/Fe] $> 0.7$ as defined in \citealt{abc+07}) after applying a carbon correction based on the evolutionary state of the star following \citet{pfb+14}.
The most metal-rich star ([Fe/H] = $-1.94$) in our sample, TucII-309, marginally misses this classification with a corrected [C/Fe] = 0.65.
For completeness, we note that pre-carbon correction, only one of our stars (TucII-305) meets the criterion for being a CEMP star.
We note that we do not resolve a gradient in the carbon abundance ([C/Fe]) as a function of distance from the center of Tucana II (see Figure~\ref{fig:spatialabundances}).
Excluding TucII-303 due to the upper limit on its [C/Fe], we derive a gradient of $0.39\pm0.20$\,dex\,deg$^-1$ which is marginally unresolved at the 2\,$\sigma$ level.
This significance is also likely over-estimated given the exclusion of TucII-303.
Consequently, we conclude that we do not find strong statistical evidence of a spatial correlation with the carbon abundance.

We derive a CEMP fraction based on the corrected carbon abundances of 75\,\% (6/8 stars) for the overall sample of stars with [Fe/H] $< -2.9$ in Tucana II.
Such a trend towards carbon-enhancement in the lowest metallicity stars is also a notable feature of the MW halo (e.g., \citealt{pfb+14} derive a halo CEMP fraction of 43\,\% after carbon correction), and suggests early enrichment by ``faint" supernovae  \citep[e.g.,][]{iut+05,nkt+13,mtk+14, sst+15} and/or spinstars {\citep[e.g.][]{meg+06, mmc+15, lsm+21}.
The prevalence of CEMP stars when [Fe/H] $< -3.0$ is generally seen in UFDs \citep[e.g.,][]{jls+20}, suggesting they may be viable progenitor environments for the CEMP stars in the Milky Way halo. 
However, differing signatures are seen in the Milky Way's larger dwarf spheroidal (dSph) galaxies \citep[e.g.,][]{sts+15, jnm+15, kgz+15, sks+17, csf+18, hem+18, chf+20}.

\subsection{$\alpha$-elements}
\label{sec:alpha}

The $\alpha$-element (Mg, Si, Ca, and Ti) abundances were derived from a variety of absorption features, the prevalence of which depended on the metallicity and {\teff} of the star and the S/N of the given spectrum. 
We note that the calcium abundance was derived from only one absorption line (4226.73\,\AA) for TucII-305, due to its low metallicity and low calcium abundance, rendering the other Ca absorption features too weak.
The Ti abundances that are plotted in Figure~\ref{fig:grid1} are the Ti II abundances, since Ti I lines are only detected in our most metal-rich star (TucII-309).
The Si abundances are generally derived from the 3905.52\,{\AA} and 4102.94\,{\AA} absorption features, which are blended with adjacent features and occupy a low S/N part of the spectra. 
Accordingly, we find that our Si abundances are not too reliable and  mark them with a colon in Table~\ref{tab:abundancetable} (see Section~\ref{sec:abundances}). 

A notable feature of $\alpha$-element abundances is a ``knee" (downturn from [$\alpha$/Fe] $\approx 0.4$) in plots of [$\alpha$/Fe] vs. [Fe/H] that indicates the onset of chemical enrichment by Type Ia Supernovae (SNe).
This signature occurs because Core Collapse SNe (CCSNe) dominate chemical evolution at early times and generally produce a constant [$\alpha$/Fe]; whereas
Type Ia SNe occur after a delay time and produce dominantly iron-peak elements, thereby suppressing [$\alpha$/Fe] at higher metallicities \citep{t+79, ibn+99, kcs+11, vmv+14, hst+19}.
The metallicity at which this departure occurs can trace the early star formation timescale \citep[e.g.,][]{tht+09}. 
As of now, one UFD (Segue 1) exhibits a flat [$\alpha$/Fe] $\approx 0.4$ across its entire metallicity range \citep[e.g.][]{fsk+14} suggesting ``one-shot" enrichment and no later star formation \citep{fb+12}.

Variations in the abundances of individual $\alpha$-elements relative to each other are sensitive to the mass of the CCSNe progenitor. 
For instance, higher mass progenitor SNe ($\gtrsim20\,M_\odot$) can lead to high [Mg/Ca] \citep[e.g.,][]{frc+04, kmg+08} whereas lower mass progenitors (e.g., $\sim10\,M_\odot$ to $\sim20\,M_\odot$) suppress the [Mg/Ca] \citep[e.g.][]{mwm+13}. 
[Mg/Ca] variations have been seen in UFDs \citep[e.g.][]{kmg+08}, in particular those associated with the LMC \citep{jls+20}, and extremely metal-poor dSph stars \citep[e.g.][]{sjf+15}. 

In our most metal-poor star (TucII-305), we find some evidence of a low [Ca/Fe] = $-0.25 \pm 0.42$, although with large uncertainty.
We note that while this abundance was derived from just one Ca line (4226.73\,{\AA}) with known NLTE effects \citep[e.g.,][]{sjf+15}, applying an NLTE correction from \citet{msp+16}\footnote{http://spectrum.inasan.ru/nLTE/} results in a negligible correction of $\sim$+0.05\,dex.
In our most metal-rich star (TucII-309), we detect a low [Mg/Ca] =  $-0.72 \pm 0.43$.
However, we argue in Sections~\ref{sec:sfh} and~\ref{sec:subch} that the signature in TucII-309 is likely due to enrichment by Type Ia SNe.
We find additional evidence of Type Ia SNe enrichment in TucII-303, due to its overall lower $\alpha$-element abundances ([Ca/Fe] = $0.09\pm0.25$, [Mg/Fe] = $0.14\pm0.38$, [Ti/Fe] = $-0.26\pm0.32$).
A full discussion of the implications of these signatures is presented in Section~\ref{sec:chemicalevolution}.
We note that there is no statistical evidence of a gradient in [Mg/Fe] (proxy for $\alpha$-elements) as a function of distance from the center of Tucana II (0.08$\pm0.40$\,dex\,deg$^{-1}$; see Figure~\ref{fig:spatialabundances}).


\subsection{Neutron-capture elements}
\label{sec:ncap}

We derived neutron-capture element (Ba, Sr, Eu) abundances through spectral syntheses.
For Sr, we synthesized the absorption feature at 4215.5\,\AA; for Ba, we synthesized the absorption feature at 4554\,\AA; and for Eu, we synthesized the absorption feature at 4129\,\AA.
We note that we do not detect Ba or Eu absorption lines in any star in our sample, and so only present upper limits on the Ba and Eu abundances.

A distinctive signature of UFD stars is a deficiency in neutron-capture abundances relative to the MW halo \citep[e.g.][]{jsf+19} although stars in some systems, e.g., Ret II and Tuc III, show a strong over-enhancement indicating enrichment by a rare $r$-process nucleosynthetic event \citep{jfc+16, rmb+16, mhs+19}. 
All of the stars in our Tucana II sample are deficient in neutron-capture abundances (see bottom panels of Figure~\ref{fig:grid2}), thus strongly supporting their association with Tucana II, despite their large distances from the center.

\subsection{Odd-Z and iron-peak elements}
\label{sec:others}

We derived abundances for the iron-peak (Cr, Mn, Co, Ni) and odd-Z elements (Na, Al, Sc) generally from multiple features.
Co and Ni absorption features were only detected in a few stars in our sample.  
Mn abundances were derived from the Mn triplet at $\sim4030$\,{\AA} for all stars.
Al abundances, similar to Si abundances in Section~\ref{sec:alpha}, are denoted with colons in Table~\ref{tab:abundancetable} as the Al absorption features in our spectra are either blended and/or have low S/N.

The iron-peak and odd-Z abundances in our sample largely track the UFD and MW halo trends in Figure~\ref{fig:grid1} and~\ref{fig:grid2}, with a few notable exceptions. 
First, Mn and Sc in TucII-309, the most metal-rich star ([Fe/H] $= -1.94$) in our sample, are strikingly deficient ([Mn/Fe] = $-1.45 \pm 0.33$ and [Sc/Fe] $= -1.29 \pm 0.48$). 
Sample syntheses of the Mn absorption line at 4030.7\,{\AA} for TucII-309 are shown in Figure~\ref{fig:MnSc}; {while we note the uncertainty on the Mn abundance is large (as seen in the syntheses), the comparison of the Mn 4754\,{\AA} absorption region in TucII-309 to that in 2MASS\,J00011195+0321051 (see bottom panel of Figure~\ref{fig:MnSc}) clearly demonstrates that TucII-309 is deficient in Mn.
A discussion of NLTE effects on the Mn abundance is presented in the last paragraph of Section~\ref{sec:subch}.
Additionally, the Sc abundance in TucII-303 ([Fe/H] $= -2.74$) is deficient also ([Sc/Fe] = $-0.82\pm0.54$).
We further discuss the implications of the low Mn of TucII-309 being a signature of a sub-Chandrasekhar mass Type Ia SN in Section~\ref{sec:discussion}.
Sc has significant uncertainties when modeling its nucleosynthetic production, so we refrain from interpreting it further in this paper. 

\section{Discussion}
\label{sec:discussion}

In this Section, we quantitatively analyze and interpret the radial velocities and detailed chemical abundances of Tucana II stars to comment on its early evolution. 
In Section~\ref{sec:dynamics}, we investigate again whether there is any evidence that the UFD is tidally disrupting and derive a dynamical mass and mass-to-light ratio. 
In Section~\ref{sec:chemicalevolution}, we use the full suite of chemical abundances of the Tucana II stars to constrain the properties of early supernovae in the system, determinate whether the star formation was extended, and discuss the signature of a sub-Chandrasekhar mass Type Ia SN in the most metal-rich star.
In Section~\ref{sec:outskirts}, we synthesize our findings to comment on the origin of the extended stellar population in Tucana II.

\begin{figure*}[!htbp]
\centering
\includegraphics[width =\textwidth]{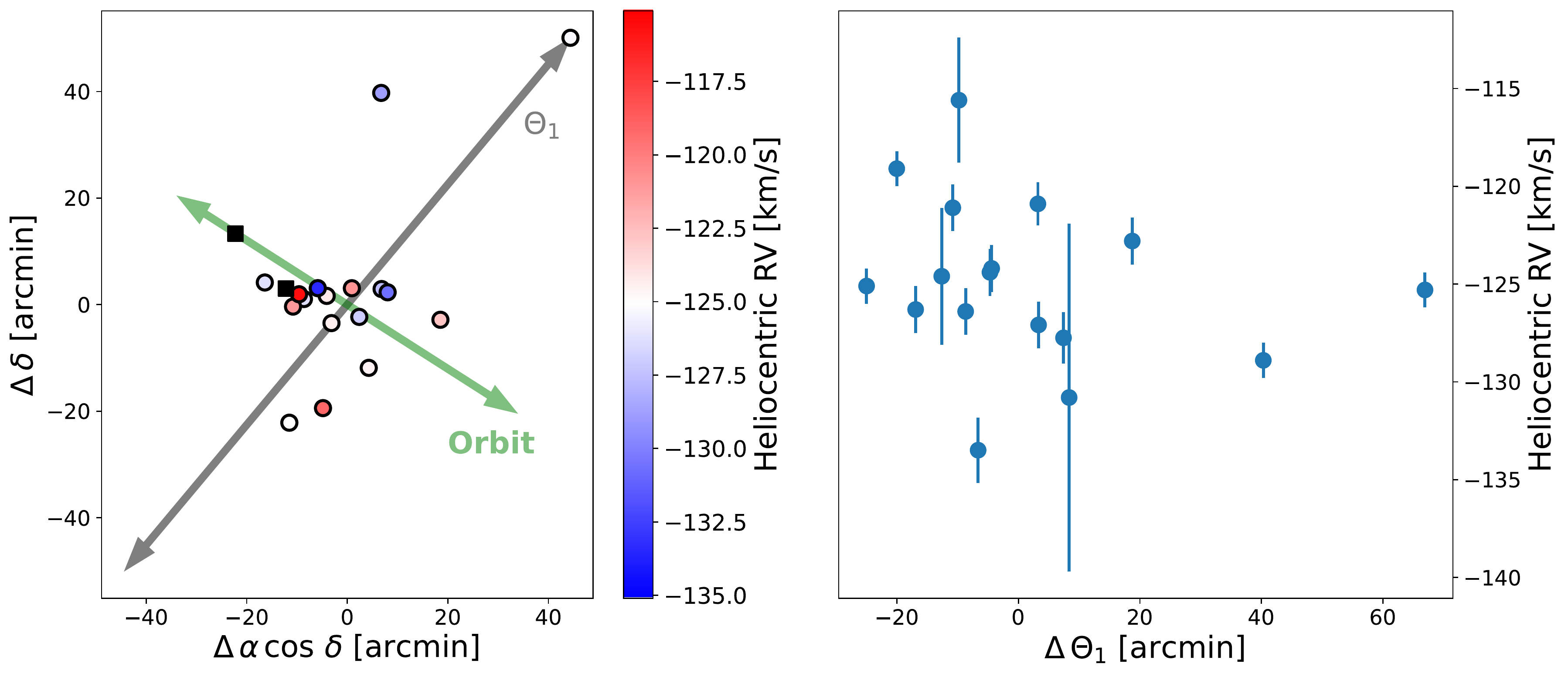}
\caption{Left: Spatial location of members of Tucana II with respect to the center of Tucana II. 
Stars are colored by their heliocentric radial velocity measurements, as presented in \citet{wmo+16, cfs+21} and this study; stars that show evidence of binarity (TucII-078 and TucII-309; see Section~\ref{sec:velocities}) are displayed as black squares. 
The arrow corresponds to the axis defined by the angle ($\theta_{1}$) along which the most distant member lies. 
The green arrow denotes the direction of Tucana II orbit from \citet{cfs+21}.
Right: Heliocentric radial velocities of Tucana II members as a function of projected distance along the axis on which the most distant member lies. 
We find no evidence of a radial velocity gradient along this axis, deriving a slope of $\text{d}v_{\text{helio}}/\text{d}\theta_1 = -2.5^{+2.9}_{-2.9}$\,km\,s$^{-1}$\,deg$^{-1}$.}
\label{fig:rvs}
\end{figure*}

\subsection{Dynamical analysis of Tucana II}
\label{sec:dynamics}

We use the radial velocities of stars in Tucana II to constrain two quantities of interest: the presence (or lack) of a radial velocity gradient, and the dynamical mass of the system within the half-light radius. 
We outlined the sample used in this analysis in the last paragraph of Section~\ref{sec:velocities}.
Determining whether the system exhibits a radial velocity gradient is particularly important since it is a key signature of a tidally disrupting dwarf galaxy \citep[e.g.][]{lsk+18}.
Consequently, a detection of a radial velocity gradient in Tucana II would readily explain its spatially extended population, due to tidal disruption.
A velocity gradient would also preclude any dynamical mass estimation of the galaxy, since it would imply that the system is not in dynamical equilibrium.

We first searched for a radial velocity gradient along the axis in the direction of the most distant star, which we denote as $\theta_1$ (see left panel of Figure~\ref{fig:rvs}).
A plot of the radial velocities of the stars as a function of projected location along this axis clearly shows no evidence of a radial velocity gradient (see right panel of Figure~\ref{fig:rvs}). 
To quantify this non-detection, we performed a Markov-Chain Monte Carlo (MCMC) analysis by implementing the maximum likelihood function in Equations 2 and 4 of \citet{wmo+15}, which includes free parameters for the mean velocity of the system, the velocity dispersion, and a velocity gradient along an arbitrary axis. 
We fix the axis to lie along $\theta_1$ for this estimator, and implement the likelihood function exactly following \citet{cfs+21} using the \texttt{emcee} python package \citep{emcee1, emcee2} with 200 walkers and 2000 steps. 
We derive a systemic velocity of the system of $\mu_{\text{rv}} = -124.7 \pm 1.0$\,km\,s$^{-1}$, a velocity dispersion of $\sigma_{\text{rv}} = 3.8^{+1.1}_{-0.7}$\,km\,s$^{-1}$, and a velocity gradient of $\text{d}v_{\text{helio}}/\text{d}\theta_1$ = $-2.5^{+2.9}_{-2.9}$\,km\,s$^{-1}$\,deg$^{-1}$, which corresponds to a non-detection of a gradient. 
We note that removing the star at $\Delta \Theta_1 \sim -0.08$, $v_{\text{helio}} \approx -134$\,km\,s$^{-1}$, which visually appears to anchor the lack of a gradient in the right panel of Figure~\ref{fig:rvs}, and redo-ing the analysis still leads to no gradient at the 2$\sigma$ level ($\text{d}v_{\text{helio}}/\text{d}\theta_1$ = $-2.8^{+2.5}_{-2.3}$\,km\,s$^{-1}$\,deg$^{-1}$).
Searching for velocity gradients around an arbitrary axis also leads to a lack of a detection ($\text{d}v_{\text{helio}}/\text{d}\theta$ = $1.8^{+6.9}_{-5.8}$\,km\,s$^{-1}$\,deg$^{-1}$).

We convert our velocity gradient to physical units (\,km\,s$^{-1}$\,kpc$^{-1}$) using the Tucana II distance from \citet{vmw+20} and find that the 95\% confidence interval on the velocity gradient along the axis of the most distant star is $-8.7$\,km\,s$^{-1}$\,kpc$^{-1}$ to $3.6$\,km\,s$^{-1}$\,kpc$^{-1}$.
This confidence interval rules out a gradient at the level that is seen in Tucana III \citep[18.3 $\pm$ 0.9\,km\,s$^{-1}$\,kpc$^{-1}$][]{lsk+18}, the only confirmed tidally disrupting UFD.
Consequently, there continues to be a lack of evidence from a dynamical analysis that the spatially extended feature in Tucana II is due to the system being tidally disrupted, as previously discussed in \citet{cfs+21}.

With the system not being significantly dynamically perturbed, we are able to derive a dynamical mass within a half-light radius of $M_{1/2}\, (r_h) = 1.6^{+1.1}_{-0.7}\times 10^6$\,M$_{\odot}$ for Tucana II, using our velocity dispersion of $3.8^{+1.1}_{-0.7}$\,km\,s$^{-1}$, a half-light radius of 120\,pc $\pm$ 30\,pc from \citet{bdb+15}, and the dynamical mass estimator in \citet{wmb+10}.
This results in a dynamical mass-to-light ratio within a half-light radius of 1020$^{+780}_{-450}$\,M$_{\odot}$/L$_{\odot}$. 
These results are consistent with the dynamical mass of $M_{1/2}\, (r_h) = 2.4^{+1.9}_{-1.2}\times 10^6$\,M$_{\odot}$ and the overall mass modeling presented in \citet{cfs+21}, which re-affirms that Tucana II is a canonical, dark-matter-dominated UFD.

For completeness, we note that limiting our sample to just stars with MIKE-based velocities and repeating the above analysis leads to a dispersion of $2.8^{+1.2}_{-0.7}$\,km\,s$^{-1}$ and a velocity gradient of $\text{d}v_{\text{helio}}/\text{d}\theta_1$ = $-2.5^{+2.5}_{-2.3}$\,km\,s$^{-1}$\,deg$^{-1}$. 
These values are still consistent with a significant dynamical mass-to-light ratio of $\sim570^{+590}_{-280}$\,M$_{\odot}$/L$_{\odot}$ (lower bound on the 95\,\% confidence interval of 150\,M$_{\odot}$/L$_{\odot}$) and a non-detection of a velocity gradient.
Accordingly, none of the interpretation changes if one chooses to minimize systematics by limiting velocity measurements to coming from a single instrument.

\subsection{The early evolution of Tucana II}
\label{sec:chemicalevolution}

In this Section, we discuss three questions related to the early evolution of Tucana II that can be addressed by the detailed chemical abundances of its member stars:
(1) The preferred properties (e.g. mass, energy) of the early supernovae in the system; 
(2) Whether Tucana II experienced relatively extended star formation; and 
(3) Whether a sub-Chandrasekhar mass Type Ia SN occurred in the system. 
We also compare each of these properties between the inner ($<$ 20\,arcmin) and outer ($>$ 20\,arcmin) stars to investigate whether these populations evolved concurrently.

\subsubsection{The early enrichment of Tucana II}\label{sec:enrichment}

Here we discuss fits from individual supernova yield models to the chemical abundances of TucII-301 and TucII-305 investigated in this study, as well as TucII-206, taken from \citet{cfj+18}. 
These stars have low enough iron abundances ([Fe/H] $< -3.3$) that their enrichment could plausibly have originated from just one individual supernova. 
We note that TucII-206 is the most metal-poor star located closer to the central population of Tucana II, whereas TucII-301 and TucII-305 are both located $>$20\,arcmin from the center of the system. 
The assumption of a single supernova enrichment event is questionable given that the metallicities of these stars are still consistent with the possibility of enrichment from multiple SNe \citep[e.g.][]{hym+18}.
However, we present fits from individual supernova yields to our chemical abundance data as a starting point for the interpretation.
We fit our abundances with the supernova yield models taken from \citet{hw+10}. 
The calculations of the best fitting models were performed exactly following \citet{jls+20}.

The results of the fits to yields of the star TucII-305 are shown in Figure~\ref{fig:starfit}. 
The nominal best-fitting model has a low progenitor mass of $\sim$11\,$M_{\odot}$ and a low explosion energy of $E\sim0.3\times10^{51}$\,ergs.
As detailed in \citet{jls+20}, the hollow squares indicate elements that were excluded from the analysis due to highly uncertain abundances (e.g., Al and Si), significant NLTE corrections (e.g., K and Mn), or model uncertainties (e.g., Sc, Cr, Cu, and Zn).  
The filled data points indicate elements that were included in the analysis.
Before the fitting procedure, the C abundances were corrected for the evolutionary state of the star following \citet{pfb+14}. 
We find that any NLTE corrections to the Mg abundances are small ($< 0.05$\,dex) based on grids presented in \citet{obl+15, ob+16}, and thus have a negligible effect on our yield fitting.
We exclude all models that lie below the red line in the bottom right panel of Figure~\ref{fig:starfit}, since those models lie below the allowed dilution mass of the halo for a given SN explosion energy (see Section~5.2 in \citealt{jls+20} for a full discussion). 

We note that while the best-fitting progenitor mass and energy is low, strong conclusions cannot be drawn since a broad range of allowed masses and energies are consistent with the abundance pattern of TucII-305 (see the bottom panels of Figure~\ref{fig:starfit}).
All models within $2\sigma$ of the $\chi^2$ value of the best-fitting yield model are included as acceptable models (grey contours) in the top panel of Figure~\ref{fig:starfit}, and are included in the histogram and plot in the bottom panels.
Qualitatively similar best-fitting models (low mass, low explosion energy) are preferred for TucII-301 and TucII-206, although their abundances still lead to a very broad range of acceptable SN progenitor models.

We provide a check on the preferred low mass and low energy progenitor CCSNe suggested by the yield modeling through the level of carbon-enhancement in the most metal-poor stars in Tucana II. 
As outlined in Section~\ref{sec:carbon}, a formation channel of carbon-enhanced metal-poor (CEMP; [C/Fe] $> 0.7$) stars are faint, mixing and fallback supernovae \citep[e.g.,][]{iut+05,nkt+13,mtk+14}.
Six out of the eight most metal-poor stars ([Fe/H] $< -2.9$) in Tucana II are CEMP stars, corroborating that faint supernovae may have dominated the early evolution of this galaxy.

\begin{figure}[!htbp]
\centering
\includegraphics[width =\columnwidth]{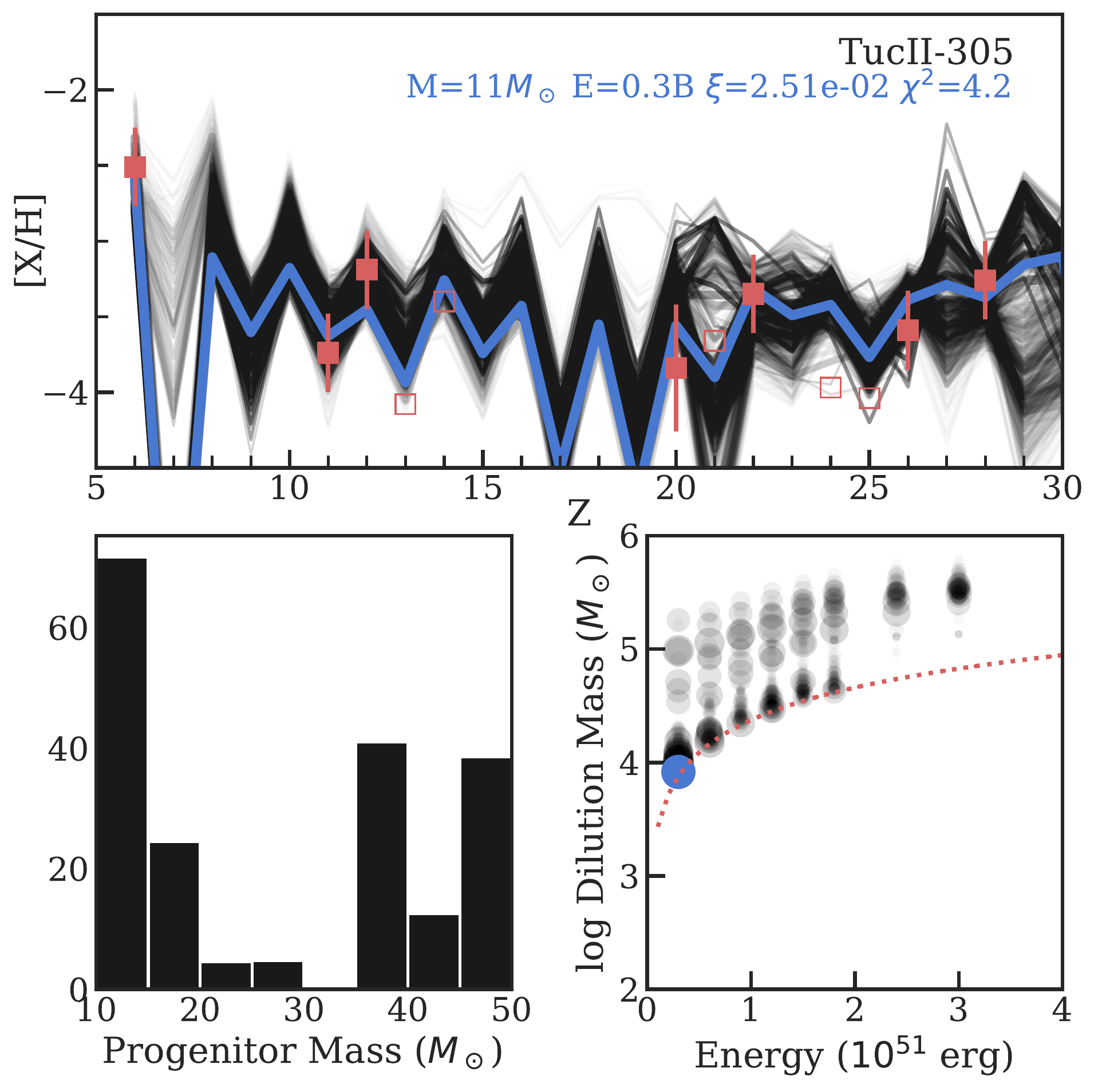}
\caption{Top panel: Abundance pattern of TucII-305 (red squares) with the best fitting supernova yield model (blue) and other supernova yield models within 2$\sigma$ of the  best-fitting $\chi^2$ value (grey) over-plotted. 
The best-fitting progenitor supernovae is low mass ($\sim11\,M_{\odot}$) and low energy (E$\sim 0.3 \times 10^{51}$\,ergs), though the abundances are consistent with a broad range of acceptable parameters.
Bottom left panel: Distribution of the progenitor masses of consistent models. While there is a preference for low masses, the range of acceptable masses does extend to the high-mass regime. 
Bottom right panel: Dilution mass vs. explosion energy of the models with the best-fitting model shown in blue. 
The red dashed line indicates the minimum dilution mass for the given supernova energy (see Section 5.2 in \citealt{jls+20}); models below this line are excluded.}
\label{fig:starfit}
\end{figure}

\subsubsection{Did Tucana II experience extended star formation?}
\label{sec:sfh}

We investigate the question of  whether Tucana II experienced extended star formation by investigating the [$\alpha$/Fe] trend in its more metal-rich stars.
As discussed in Section~\ref{sec:alpha} and recent work in the literature \citep[e.g.][]{kxg+19}, a declining [$\alpha$/Fe] ratio at higher metallicities suggests enrichment by Type Ia supernovae (SNe).
This is because the timescale for the onset of enrichment by Type Ia supernovae is set by their delay time distribution. 
While the exact minimum delay time of Type Ia SNe remains somewhat ambiguous \citep[e.g.][]{mg+17}, a star with evidence of strong enrichment by Type Ia SNe still likely formed at a later time than the predominantly CCSNe-enriched stars that formed in an initial period of star formation.

Interestingly, we see evidence of a declining [$\alpha$/Fe] trend in the most metal-rich star in the central population (TucII-033; [Fe/H] = $-2.41 \pm 0.12$ and [$\alpha$/Fe] = $0.07 \pm 0.14$), and also the two most metal-rich stars in the outskirts, TucII-303 ([Fe/H] = $-2.74 \pm 0.17$; [$\alpha$/Fe] = $-0.06 \pm 0.12$) and TucII-309 ([Fe/H] = $-1.94\pm0.22$; [$\alpha$/Fe] = $0.02 \pm 0.14$).
This would naively suggest that these stars reflect enrichment by Type Ia SNe. 
As such, the metallicity spread in Tucana II is due to, at least in some part, chemical evolution stretching past the formation of its most metal-poor stars (at [Fe/H] $\lesssim -2.9$) for stars located both in the inner regions and the outskirts.

However, this picture is slightly complicated by the most metal-rich star (TucII-309) due to its apparently high calcium abundance ([Ca/Fe] = $0.38 \pm 0.24$).
This abundance appears to conflict with the downward trend of the other $\alpha$-element abundances ([Mg/Fe] = $-0.34 \pm 0.25$; [Ti/Fe] = $0.00 \pm 0.25$).
To determine whether this Ca abundance is in tension with enrichment by Type Ia SNe, we derive the fraction of material from Type Ia SNe ($f_{\text{Ia}}$) that enriched this star as indicated by its Ca abundance and accompanying uncertainty. 
We do this analysis by adopting a CCSN yield of [Ca/Fe] = 0.32 (the average abundance of stars with [Fe/H] $< -2.9$ in Tucana II) and adopting Type Ia SN yields from Table~1 in \citet{kxg+19}.
We derive an upper limit on the 95\,\% confidence interval of $f_{\text{Ia}} \lesssim$ 0.85 from [Ca/Fe] $> -0.09$ (the lower bound on the 95\% confidence interval from [Ca/Fe] = $0.38 \pm 0.24$), fairly independent of the various [Ca/Fe] yields of Type Ia SNe presented in \citet{kxg+19}.
This high upper limit on $f_{\text{Ia}}$ indicates that the [Ca/Fe] in TucII-309 does not exclude significant contribution from Type Ia SNe.

Performing the same exercise for Mg in TucII-309 returns $f_{\text{Ia}} = 0.85^{+0.07}_{-0.13}$ with a lower limit on the 95\,\% confidence interval of $f_{\text{Ia}} \gtrsim 0.5$ (from [Mg/Fe] $< 0.15$).
This tighter constraint on $f_{\text{Ia}}$ occurs because SNe Ia produce negligible Mg but a non-negligible Ca ([Ca/Fe] $\sim -0.25$), making the Mg abundance a more sensitive discriminator of the contribution from SNe Ia \citep[e.g., Table 1 in][]{kxg+19}. 
Combining the results from the Ca and Mg abundances supports a significant contribution of SNe Ia material ($0.5 \lesssim f_{\text{Ia}} \lesssim 0.85$) enriching TucII-309. 
Therefore, star formation in Tucana II lasted long enough for SNe Ia to impact the chemical abundances of its stars.

\subsubsection{A sub-Chandrasekhar mass Type Ia supernova in Tucana II?}
\label{sec:subch}

The low Mn abundance ([Mn/Fe] = $-1.38 \pm 0.33$) of TucII-309, coupled with its low $\alpha$-abundances (see Section~\ref{sec:sfh}), indicates enrichment by low metallicity sub-Chandrasekhar mass (sub-$M_{\text{Ch}}$) Type Ia SNe.
Mn deficiency is a key characteristic of yields from sub-$M_{\text{Ch}}$ Type Ia SNe distinguishing them from yields of a $M_{\text{Ch}}$ Type Ia SN (see discussion in \citealt{mpb+18}). 
Mn abundance deficiencies in other dwarf galaxies have been used to gauge the contribution of the sub-$M_{\text{Ch}}$ Type Ia SNe channel in recent works \citep[e.g.,][]{kxg+19, dks+20}.

However, what is particularly striking about the Mn abundance in TucII-309 is its extreme deficiency: no other UFD star in our literature compilation has a [Mn/Fe] $< -1.0$ (see top left panel of Figure~\ref{fig:grid2}), and only two halo stars, HE~1310$-$0536 \citep{hhc+14} and HE~2215$-$2548 \citep{cct+13}, have lower [Mn/Fe] in the JINAbase\footnote{https://jinabase.pythonanywhere.com} compilation of metal-poor stars \citep{af+18}. 
The compilation of yields from sub-$M_{\text{Ch}}$ Type Ia models in Figure~6 of \citet{dks+20} show yields ranging from [Mn/Fe] $\approx -2.0$ to $\approx -0.5$ at [Fe/H] $\approx -2.0$ \citep{skm+18, bbm+19, ln+20}.
Thus, based on the [Mn/Fe] = $-1.38 \pm 0.33$ value in TucII-309,  it is plausible that the star formed in a region that was very dominantly enriched by a sub-$M_{\text{Ch}}$ Type Ia.
The yield of this Type Ia SN was then likely not homogeneously mixed into the full UFD halo, given that efficient mixing likely would remove such a prominent [Mn/Fe] deficiency.
The [Mn/Fe] deficiency in TucII-309 can also be viewed as evidence that it has significant enrichment by material from SNe Ia \citep[e.g.,][]{mpb+18}.

For completeness, we note that a NLTE correction on [Mn/Fe] for a red giant branch star at the metallicity of TucII-309 is likely $\lesssim 0.37$\,dex (following Eq. 10 from \citealt{dks+20} based on Figure~9 in \citealt{mge+19}).
The [Mn/Fe] correction generally increases at lower metallicities.
This means that applying a NLTE correction to the Mn abundances of our sample would likely make the Mn deficiency in TucII-309 stand out more sharply, since it is the most metal-rich star.

\subsection{The formation of the outskirts of Tucana II}
\label{sec:outskirts}

\citet{cfs+21} discuss three scenarios for the formation of the outskirts of Tucana II: tidal disruption of the system, in-situ formation through supernova feedback, or a merger between two (or more) building block galaxies. 
In this subsection, we re-visit each of these scenarios in light of our results presented in Section~\ref{sec:analysis} and the discussion in Sections~\ref{sec:dynamics} and~\ref{sec:chemicalevolution}. 

\textit{Tidal disruption} -- \citet{cfs+21} present two lines of evidence disfavoring tidal disruption as a formation channel for the outskirts of Tucana II. 
The first is that the location of predicted tidal debris is misaligned with the location of the spatially distant stars in the system (see Figure~1 in \citealt{cfs+21}). 
The second is the lack of a detected velocity gradient in the system, which is a known signature of tidally disrupting dwarf galaxies \citep[e.g.][]{lsk+18}. 
The upper limit of the velocity gradient from \citet{cfs+21} is 27\,km\,s$^{-1}$\,deg$^{-1}$.

Our analysis in Section~\ref{sec:dynamics} strongly re-affirms the lack of a detected velocity gradient, with a derived value of $\text{d}v_{\text{helio}}/\text{d}\theta_1$ = $-2.5^{+2.9}_{-2.9}$\,km\,s$^{-1}$\,deg$^{-1}$ and a 95\% confidence interval ranging from $-8.7$\,km\,s$^{-1}$\,kpc$^{-1}$ to $3.6$\,km\,s$^{-1}$\,kpc$^{-1}$.
This stronger limit largely follows from the increased velocity precision of our MIKE spectra compared to the previous MagE spectra. 
Notably, our new confidence interval on the velocity gradient of Tucana II excludes a velocity gradient at the level of Tucana III ($18.3^{+0.9}_{-0.9}$\,km\,s$^{-1}$\,kpc$^{-1}$; \citealt{lsk+18}), the only confirmed tidally disrupting dwarf galaxy.

However, we note that it remains somewhat unclear at what level a radial velocity gradient needs to be excluded to conclusively state that the system is not tidally disrupting. 
A pre-requisite to derive a theoretical predicted velocity gradient would be to generate a tidal disruption model with debris that reproduces the direction of the extended feature in Tucana II.
Unfortunately, the spatial location of the outlying stars is incompatible with a standard tidal disruption model \citep{cfs+21}. 
More extensive modeling is beyond the scope of this paper.

\textit{Supernova feedback} -- \citet{cfs+21} suggest early, energetic supernovae (SNe) or bursty feedback as one possible scenario for ``puffing up" Tucana II, leading to a spatially extended stellar feature. 
There is some theoretical motivation that UFDs may have experienced early, bursty star formation \citep[e.g.,][]{whp+19}. 

We do not find evidence that Tucana II experienced particularly energetic SNe, based on the chemical abundances of its member stars as detailed below.
The most metal-poor stars in the system do not prefer enrichment by energetic supernovae (Section~\ref{sec:enrichment}). 
And the more metal-rich stars have abundance trends that do not generally deviate from the overall UFD trend (Figures~\ref{fig:grid1} and~\ref{fig:grid2}).
This lack of deviation suggests that Tucana II was enriched by CC SNe that are not particularly different (at least, in terms of their yields and likely energies) from those that occurred in other UFDs. 
We do note that the only anomalous abundance signature in Tucana II is the evidence for a sub-Chandrasekhar mass Type Ia SN dominating the (later time) enrichment of TucII-309 (see Section~\ref{sec:subch}) and a Sc under-abundance in the most metal-rich star.
However, such sub-$M_{\text{Ch}}$ Ia SNe are not generally more energetic than other Type Ia SNe \cite[e.g.][]{bbm+19}, and the production of Sc has meaningful modeling discrepancies with observed signatures \citep[e.g.,][]{kcu+06, cs+15}.
More generally, UFDs likely do not experience sufficient feedback to core their dark matter halos (see Figure~13 in \citealt{bb+17}, based on simulations in \citealt{tmd+16, fbe+17, hwk+18, ckw+18}), making it questionable whether they would sufficiently ``puff up" their stellar component too.

Consequently, our sample of chemical abundances does not especially favor this formation channel.
Since there is no evidence that Tucana II hosted particularly distinct SN types from other UFDs, if this formation channel is the explanation for its extended stellar component, then most other UFDs should show similar features as well.
Otherwise, if Tucana II is shown to have experienced particularly bursty star formation relative to other UFDs, then this scenario remains plausible. 
Future work on detecting extended features around other UFDs, as well as with any additional investigations of the star formation history of Tucana II, is needed to further constrain the validity of this channel. 

\textit{Merging of First Galaxies} -- Recent work on modeling Tucana II by \citet{tyf+21} has affirmed that spatially extended features around UFDs can be formed by early galactic mergers. 
Specifically, \citet{tyf+21} find that an early galactic merger between two first galaxies, several $100$\,Myr after formation, leads to a spatially extended feature with a very weak metallicity gradient (see Figure 4 in \citealt{tyf+21}). 
The merger deposits the stars of the colliding galaxies in the outskirts of the final, merged galaxy, and star formation is triggered in the central region of the system.
Stars formed during or after the merger can still be located at large distances from the center of the galaxy (e.g., Figure 4 in \citealt{tyf+21}), so the declining [$\alpha$/Fe] seen in the inner and outer stars of Tucana II is still consistent with this interpretation.

Such a formation scenario, which is concordant with a delayed episode of star formation, is consistent with our chemical abundance data.
The ``knee"  in the $\alpha$-element abundances seen in the Tucana II sample favors some extended/delayed star formation in the system (see Sections~\ref{sec:alpha} and~\ref{sec:sfh}). 

Generalizing this result across UFDs, we might expect that UFDs without an $\alpha$-element ``knee" in their abundance trends \citep[e.g., Segue 1; ][]{fsk+14} are less likely to be assembled systems and thus less likely to show these extended features.
This picture is consistent with surviving first galaxies not showing a decline in $\alpha$-element abundances, such as Segue 1, as described in the one-shot enrichment scenario of \citet{fb+12}.

\section{Conclusion}
\label{sec:conclusion}

In this paper, we present a dynamical and chemical study of stars in the outskirts (0.3\,kpc to 1.1\,kpc) of the Tucana II UFD based on high-resolution ($R\sim25,000$) Magellan/MIKE spectra.
We derive low metallicities ($-3.6 <$ [Fe/H] $< -1.9$), low surface gravities ($\logg < 2.0$), and low neutron-capture element abundances ([Ba/Fe] $< -0.7$; [Sr/Fe] $\leq -0.8$) for these stars, affirming them as members of Tucana II despite their large distances from the center of the UFD.

We do not detect a radial velocity gradient in Tucana II, despite our expanded sample of members and the precise ($\sim$1\,km\,s$^{-1}$ uncertainty) velocities that were derived from our MIKE spectra. 
We find that the 95\% confidence interval of the velocity gradient ranges from $-8.7$\,km\,s$^{-1}$\,kpc$^{-1}$ to $3.6$\,km\,s$^{-1}$\,kpc$^{-1}$ (Section~\ref{sec:dynamics}), a far more stringent constraint than the previous limit of $\lesssim 27$\,km\,s$^{-1}$\,kpc$^{-1}$ in \citet{cfs+21}.
This lack of a velocity gradient does not lend support to the extended nature of the system being due to significant tidal disruption.
We derive a dynamical mass of $M_{1/2}\, (r_h) = 1.6^{+1.1}_{-0.7}\times 10^6$\,M$_{\odot}$ within the half-light radius and a corresponding mass-to-light ratio of $\sim1020^{+780}_{-450}\,$M$_{\odot}$/L$_{\odot}$.

The detailed chemical abundances of stars in Tucana II are largely similar to what is found in other UFD stars (see Figures~\ref{fig:grid1} and~\ref{fig:grid2}).
We derive a large fraction (75\,\%) of carbon-enhanced metal-poor stars below [Fe/H] $= -2.9$, suggesting that faint SNe may have dominated the early enrichment of Tucana II.
This picture is consistent with fitting individual SNe yield models to the chemical abundance pattern of the most metal-poor star in the UFD (Section~\ref{sec:enrichment}). 
We find evidence for extended/delayed star formation due to the downturn in [$\alpha$/Fe] in the more metal-rich stars ([Fe/H] $\gtrsim -2.8$) in Tucana II (Section~\ref{sec:sfh}), and for localized, heavy enrichment by a sub-Chandrasekhar mass Type Ia SN in the most metal-rich star due to its extremely low [Mn/Fe] = $-1.38 \pm 0.33$ (Section~\ref{sec:subch}).

We re-evaluate the formation channels of the outlying regions of Tucana II that were discussed in \citet{cfs+21}, in light of our new data (see Section~\ref{sec:outskirts} for a full discussion).
We still find no evidence for tidal disruption due to the lack of a detected velocity gradient. 
The general consistency between the chemical abundances of stars in Tucana II and other UFDs suggests that the SNe in Tucana II were not particularly energetic.
This result disfavors SNe injecting energy and causing Tucana II to ``puff up" to form its outlying regions, unless the system experienced particularly bursty star formation or most other UFDs also show such extended features, since there is no evidence that SNe in Tucana II were unusually energetic.

Our evidence for delayed/extended star formation in Tucana II (from its $\alpha$-element abundances) is qualitatively consistent with the outlying regions being formed by an early merger between two first galaxies \citep{tyf+21}.
However, the unconstrained timescales of the star formation in Tucana II prevent a quantitative comparison.
We hypothesize that if early galactic mergers as outlined in \citet{tyf+21} were the dominant formation channel for outlying regions in UFDs, then UFDs with no  $\alpha$-element ``knees" (e.g., Segue 1) should be less likely to show extended stellar features.

Future work on detecting extended features around other dwarf galaxies is already underway \citep{lja+21, fw+21, qzp+21, yhj+22}.
As shown in this paper, population-level insights on the dynamics and detailed chemical abundances of the stars in these extended features can disentangle the processes that govern early galaxy formation, assembly, and evolution.  
Such work is currently restricted to the brightest stars in these systems, but is especially suited for spectroscopy with upcoming thirty meter-class telescopes.\\\\


A.C. is supported by a Brinson Prize Fellowship at the University of Chicago/KICP. 
A.C. and A.F. acknowledge support from NSF grant AST-1716251. 
We thank Kaitlin Rasmussen for collecting the December MIKE data for TucII-303.
We thank an anonymous referee for improving the content of the paper.
This work made use of NASA's Astrophysics Data System Bibliographic Services, and the SIMBAD database, operated at CDS, Strasbourg, France \citep{woe+00}.

The national facility capability for SkyMapper has been funded through ARC LIEF grant LE130100104 from the Australian Research Council, awarded to the University of Sydney, the Australian National University, Swinburne University of Technology, the University of Queensland, the University of Western Australia, the University of Melbourne, Curtin University of Technology, Monash University and the Australian Astronomical Observatory. SkyMapper is owned and operated by The Australian National University's Research School of Astronomy and Astrophysics.

This work has made use of data from the European Space Agency (ESA) mission
{\it Gaia} (\url{https://www.cosmos.esa.int/gaia}), processed by the {\it Gaia}
Data Processing and Analysis Consortium (DPAC,
\url{https://www.cosmos.esa.int/web/gaia/dpac/consortium}). Funding for the DPAC
has been provided by national institutions, in particular the institutions
participating in the {\it Gaia} Multilateral Agreement.


\facilities{Magellan:Clay (MIKE), SkyMapper}

\software{Astropy \citep{astropy, astropy2}, NumPy \citep{numpy}, SciPy \citep{scipy}, Matplotlib \citep{Hunter+07}, emcee \citep{emcee1, emcee2}, SMH \citep{c+14}, MOOG \citep{s+73}, alexmods (https://github.com/alexji/alexmods), MIKE CarPy reduction pipeline \citep{k+03}.}

\bibliography{skymapper}

\startlongtable
\begin{deluxetable*}{lrrrrr|@{\hskip 0.1in}lrrrrrr} 
\tablecaption{$\sigma$ values correspond to random uncertainties. 
See Table~\ref{tab:uncertainties} for total uncertainties.
Colons indicate measurements with large uncertainties.}
\tablecolumns{12}
\tablewidth{\textwidth}
\tablecaption{Chemical abundances\label{tab:abundancetable}}
\tablehead{   
  \colhead{El.} &
  \colhead{N} &
  \colhead{$\log\epsilon(\text{X})_{\odot}$} &
  \colhead{[X/H]} & 
  \colhead{[X/Fe]} & 
  \colhead{$\sigma\tablenotemark{a}$} {\hskip 0.05in} &
  \colhead{El.} &
  \colhead{N} &
  \colhead{$\log\epsilon(\text{X})_{\odot}$} &
  \colhead{[X/H]} & 
  \colhead{[X/Fe]} & 
  \colhead{$\sigma\tablenotemark{a}$} 
}
\startdata
\multicolumn{6}{c}{\textbf{TucII-301}} &
\multicolumn{6}{c}{\textbf{TucII-303}}\\
CH  &  2  & 8.43 & $-$2.79 & 0.62 & 0.23 {\hskip 0.05in}  & {\hskip 0.05in} CH  &  2  & 8.43 & $<-$2.14 & $<$ 0.60 & \nodata \\ 
CH\tablenotemark{b}  &  2  & 8.43 & $-$2.50 & 0.91 & 0.23 {\hskip 0.05in}  & {\hskip 0.05in} CH\tablenotemark{b}  &  2  & 8.43 & $<-$1.77 & $<$ 0.97 & \nodata \\ 
Na I  &  2  & 6.24 & $-$3.28 & 0.13 & 0.29 {\hskip 0.05in}  & {\hskip 0.05in} Na I  &  2  & 6.24 & $-$2.90 & $-$0.16 & 0.17 \\ 
Mg I  &  4  & 7.60 & $-$2.90 & 0.51 & 0.23 {\hskip 0.05in}  & {\hskip 0.05in} Mg I  &  2  & 7.60 & $-$2.60 & 0.14 & 0.17 \\ 
Al I  &  2  & 6.45 & $-$3.85:\tablenotemark{c} & $-$0.44:\tablenotemark{c} & \nodata {\hskip 0.05in}  & {\hskip 0.05in} Al I  &  1  & 6.45 & $<-$3.04 & $<-$0.30 & \nodata \\ 
Si I  &  1  & 7.51 & $-$3.01:\tablenotemark{c} & 0.40:\tablenotemark{c} & \nodata {\hskip 0.05in}  & {\hskip 0.05in} Si I  &  1  & 7.51 & $-$2.85:\tablenotemark{c} & $-$0.11:\tablenotemark{c} & \nodata \\ 
Ca I  &  3  & 6.34 & $-$3.10 & 0.31 & 0.23 {\hskip 0.05in}  & {\hskip 0.05in} Ca I  &  5  & 6.34 & $-$2.65 & 0.09 & 0.17 \\ 
Sc II  &  4  & 3.15 & $-$3.08 & 0.33 & 0.23 {\hskip 0.05in}  & {\hskip 0.05in} Sc II  &  2  & 3.15 & $-$3.56 & $-$0.82 & 0.23 \\ 
Ti II  &  10  & 4.95 & $-$2.92 & 0.49 & 0.28 {\hskip 0.05in}  & {\hskip 0.05in} Ti II  &  9  & 4.95 & $-$3.00 & $-$0.26 & 0.17 \\ 
Cr I  &  2  & 5.64 & $-$3.85 & $-$0.44 & 0.32 {\hskip 0.05in}  & {\hskip 0.05in} Cr I  &  3  & 5.64 & $-$2.99 & $-$0.25 & 0.26 \\ 
Mn I  &  2  & 5.43 & $-$4.05 & $-$0.64 & 0.23 {\hskip 0.05in}  & {\hskip 0.05in} Mn I  &  3  & 5.43 & $-$3.24 & $-$0.50 & 0.45 \\ 
Fe I  &  31  & 7.50 & $-$3.41 & 0.00 & 0.23 {\hskip 0.05in}  & {\hskip 0.05in} Fe I  &  50  & 7.50 & $-$2.74 & 0.00 & 0.17 \\ 
Fe II  &  2  & 7.50 & $-$3.40 & 0.01 & 0.23 {\hskip 0.05in}  & {\hskip 0.05in} Fe II  &  7  & 7.50 & $-$2.72 & 0.02 & 0.17 \\ 
Sr II  &  1  & 2.87 & $<-$4.56 & $<-$1.15 & \nodata {\hskip 0.05in}  & {\hskip 0.05in} Sr II  &  1  & 2.87 & $-$3.54 & $-$0.80 & 0.65 \\ 
Ba II  &  1  & 2.18 & $<-$4.11 & $<-$0.70 & \nodata {\hskip 0.05in}  & {\hskip 0.05in} Ba II  &  1  & 2.18 & $<-$4.24 & $<-1.50$ & \nodata \\ 
Eu I  &  1  & 0.52 & $<-$2.31 & $<$ 1.10 & \nodata {\hskip 0.05in}  & {\hskip 0.05in} Eu I  &  1  & 0.52 & $<-$1.64 & $<$ 1.10 & \nodata \\ 
\\
\multicolumn{6}{c}{\textbf{TucII-305}} &
\multicolumn{6}{c}{\textbf{TucII-306}}\\
CH  &  2  & 8.43 & $-$2.78 & 0.81 & 0.26 {\hskip 0.05in}  & {\hskip 0.05in} CH  &  2  & 8.43 & $-$2.74 & 0.52 & 0.22 \\ 
CH\tablenotemark{b}  &  2  & 8.43 & $-$2.51 & 1.08 & 0.26 {\hskip 0.05in}  & {\hskip 0.05in} CH\tablenotemark{b}  &  2  & 8.43 & $-$2.35 & 0.91 & 0.22 \\ 
Na I  &  2  & 6.24 & $-$3.74 & $-$0.15 & 0.26 {\hskip 0.05in}  & {\hskip 0.05in} Na I  &  2  & 6.24 & $-$2.90 & 0.36 & 0.22 \\ 
Mg I  &  3  & 7.60 & $-$3.19 & 0.40 & 0.26 {\hskip 0.05in}  & {\hskip 0.05in} Mg I  &  6  & 7.60 & $-$2.55 & 0.71 & 0.22 \\ 
Al I  &  2  & 6.45 & $-$4.08:\tablenotemark{c} & $-$0.49:\tablenotemark{c} & \nodata {\hskip 0.05in}  & {\hskip 0.05in} Al I  &  2  & 6.45 & $-$3.00:\tablenotemark{c} & 0.26:\tablenotemark{c} & \nodata \\ 
Si I  &  1  & 7.51 & $-$3.40:\tablenotemark{c} & 0.19:\tablenotemark{c} & \nodata {\hskip 0.05in}  & {\hskip 0.05in} Si I  &  2  & 7.51 & $-$2.24:\tablenotemark{c} & 1.02:\tablenotemark{c} & \nodata \\ 
Ca I  &  2  & 6.34 & $-$3.84 & $-$0.25 & 0.42 {\hskip 0.05in}  & {\hskip 0.05in} Ca I  &  4  & 6.34 & $-$2.90 & 0.36 & 0.22 \\ 
Sc II  &  4  & 3.15 & $-$3.66 & $-$0.07 & 0.26 {\hskip 0.05in}  & {\hskip 0.05in} Sc II  &  5  & 3.15 & $-$3.09 & 0.17 & 0.22 \\ 
Ti II  &  11  & 4.95 & $-$3.35 & 0.24 & 0.26 {\hskip 0.05in}  & {\hskip 0.05in} Ti II  &  15  & 4.95 & $-$2.80 & 0.46 & 0.27 \\ 
Cr I  &  2  & 5.64 & $-$3.97 & $-$0.38 & 0.26 {\hskip 0.05in}  & {\hskip 0.05in} Cr I  &  3  & 5.64 & $-$3.63 & $-$0.37 & 0.22 \\ 
Mn I  &  3  & 5.43 & $-$4.04 & $-$0.45 & 0.26 {\hskip 0.05in}  & {\hskip 0.05in} Mn I  &  2  & 5.43 & $-$3.89 & $-$0.63 & 0.22 \\ 
Fe I  &  41  & 7.50 & $-$3.59 & 0.00 & 0.26 {\hskip 0.05in}  & {\hskip 0.05in} Fe I  &  59  & 7.50 & $-$3.26 & 0.00 & 0.22 \\ 
Fe II  &  2  & 7.50 & $-$3.56 & 0.03 & 0.26 {\hskip 0.05in}  & {\hskip 0.05in} Fe II  &  4  & 7.50 & $-$3.24 & 0.02 & 0.22 \\ 
Ni I  &  1  & 6.22 & $-$3.26 & 0.33 & 0.26 {\hskip 0.05in}  & {\hskip 0.05in} Co I  &  2  & 4.99 & $-$3.22 & 0.04 & 0.42 \\ 
Sr II  &  1  & 2.87 & $<-$4.99 & $<-$1.40 & \nodata {\hskip 0.05in}  & {\hskip 0.05in} Sr II  &  1  & 2.87 & $-$4.63 & $-$1.37 & 0.30 \\ 
Ba II  &  1  & 2.18 & $<-$4.74 & $<-$1.15 & \nodata {\hskip 0.05in}  & {\hskip 0.05in} Ba II  &  1  & 2.18 & $<-$4.46 & $<-$1.20 & \nodata \\ 
Eu I  &  1  & 0.52 & $<-$2.14 & $<$ 1.45 & \nodata {\hskip 0.05in}  & {\hskip 0.05in} Eu I  &  1  & 0.52 & $<-$2.21 & $<$ 1.05 & \nodata \\ 
\\
\multicolumn{6}{c}{\textbf{TucII-309}}\\
CH  &  2  & 8.43 & $-$1.46 & 0.48 & 0.22{\hskip 0.05in} \\ 
CH\tablenotemark{b}  &  2  & 8.43 & $-$1.29 & 0.65 & 0.22{\hskip 0.05in} \\ 
Na I  &  2  & 6.24 & $-$2.44 & $-$0.50 & 0.22{\hskip 0.05in} \\ 
Mg I  &  5  & 7.60 & $-$2.28 & $-$0.34 & 0.22{\hskip 0.05in} \\ 
Al I  &  1  & 6.45 & $-$1.93:\tablenotemark{c} & 0.01:\tablenotemark{c} & \nodata{\hskip 0.05in} \\ 
Si I  &  2  & 7.51 & $-$2.52:\tablenotemark{c} & $-$0.58:\tablenotemark{c} & \nodata{\hskip 0.05in} \\ 
Ca I  &  15  & 6.34 & $-$1.56 & 0.38 & 0.22{\hskip 0.05in} \\ 
Sc II  &  2  & 3.15 & $-$3.23 & $-$1.29 & 0.33{\hskip 0.05in} \\ 
Ti I  &  6  & 4.95 & $-$2.02 & $-$0.08 & 0.22{\hskip 0.05in} \\ 
Ti II  &  18  & 4.95 & $-$1.94 & 0.00 & 0.24{\hskip 0.05in} \\ 
Cr II  &  1  & 5.64 & $-$1.50 & 0.44 & 0.40{\hskip 0.05in} \\ 
Cr I  &  13  & 5.64 & $-$1.87 & 0.07 & 0.23{\hskip 0.05in} \\ 
Mn I  &  3  & 5.43 & $-$3.32 & $-$1.38 & 0.22{\hskip 0.05in} \\ 
Fe I  &  126  & 7.50 & $-$1.94 & 0.00 & 0.22{\hskip 0.05in} \\ 
Fe II  &  10  & 7.50 & $-$1.91 & 0.03 & 0.33{\hskip 0.05in} \\ 
Sr II  &  1  & 2.87 & $-$3.73 & $-$1.79 & 0.50{\hskip 0.05in} \\ 
Ba II  &  1  & 2.18 & $<-$4.03 & $<-$2.09 & \nodata{\hskip 0.05in} \\ 
Eu I  &  1  & 0.52 & $<-$1.73 & $<$ 0.21 & \nodata{\hskip 0.05in} \\ 
\enddata
\tablenotetext{a}{Random uncertainties. See Table~\ref{tab:uncertainties} for total uncertainties.}
\tablenotetext{b}{Corrected for the evolutionary status of the star following \citet{pfb+14}.}
\tablenotetext{c}{Colons (:) indicate large uncertainties despite the detection of a line feature.}
\end{deluxetable*}

\startlongtable
\begin{deluxetable*}{lrrrrrr|lrrrrrr} 
\tabletypesize{\footnotesize}
\tablecolumns{10}
\tablecaption{Uncertainties\label{tab:uncertainties}}
\tablehead{   
 \colhead{El.} & 
 \colhead{N} &
  \colhead{$\sigma_{\text{rand}}$} &
  \colhead{$\sigma_{\text{[X/H]sys}}$} & 
  \colhead{$\sigma_{\text{[X/H]tot}}$} & 
  \colhead{$\sigma_{\text{[X/Fe]sys}}$} & 
  \colhead{$\sigma_{\text{[X/Fe]tot}}$} & 
  \colhead{El.} &
  \colhead{N} &
  \colhead{$\sigma_{\text{rand}}$} &
  \colhead{$\sigma_{\text{[X/H]sys}}$} & 
  \colhead{$\sigma_{\text{[X/H]tot}}$} &
  \colhead{$\sigma_{\text{[X/Fe]sys}}$} & 
  \colhead{$\sigma_{\text{[X/Fe]tot}}$} 
}
\startdata
\multicolumn{7}{c}{\textbf{TucII-301}} &
\multicolumn{7}{c}{\textbf{TucII-303}}\\
CH & 2 & 0.23 & 0.55 & 0.60 & 0.30 & 0.38 {\hskip 0.05in}  & {\hskip 0.05in} CH & 1 & \nodata & \nodata & \nodata & \nodata & \nodata\\ 
Na I & 2 & 0.29 & 0.26 & 0.39 & 0.13 & 0.32 {\hskip 0.05in}  & {\hskip 0.05in} Na I & 2 & 0.17 & 0.21 & 0.27 & 0.06 & 0.18 \\ 
Mg I & 4 & 0.23 & 0.35 & 0.42 & 0.14 & 0.27 {\hskip 0.05in}  & {\hskip 0.05in} Mg I & 2 & 0.17 & 0.34 & 0.38 & 0.21 & 0.27 \\ 
Al I & 2 & \nodata & \nodata & \nodata & \nodata & \nodata {\hskip 0.05in}  & {\hskip 0.05in} Al I & 1 & \nodata & \nodata & \nodata & \nodata & \nodata\\ 
Si I & 1 & \nodata & \nodata & \nodata & \nodata & \nodata {\hskip 0.05in}  & {\hskip 0.05in} Si I & 1 & \nodata & \nodata & \nodata & \nodata & \nodata\\  
Ca I & 3 & 0.23 & 0.39 & 0.45 & 0.25 & 0.34 {\hskip 0.05in}  & {\hskip 0.05in} Ca I & 5 & 0.17 & 0.19 & 0.25 & 0.11 & 0.20 \\ 
Sc II & 4 & 0.23 & 0.30 & 0.38 & 0.23 & 0.33 {\hskip 0.05in}  & {\hskip 0.05in} Sc II & 2 & 0.23 & 0.67 & 0.71 & 0.49 & 0.54 \\ 
Ti II & 10 & 0.28 & 0.24 & 0.37 & 0.11 & 0.30 {\hskip 0.05in}  & {\hskip 0.05in} Ti II & 9 & 0.17 & 0.27 & 0.32 & 0.06 & 0.18 \\ 
Cr I & 2 & 0.32 & 0.34 & 0.47 & 0.04 & 0.32 {\hskip 0.05in}  & {\hskip 0.05in} Cr I & 3 & 0.26 & 0.31 & 0.40 & 0.11 & 0.28 \\ 
Mn I & 2 & 0.23 & 0.36 & 0.43 & 0.16 & 0.28 {\hskip 0.05in}  & {\hskip 0.05in} Mn I & 3 & 0.45 & 0.35 & 0.57 & 0.25 & 0.51 \\ 
Fe I & 31 & 0.23 & 0.35 & 0.42 & \nodata & \nodata {\hskip 0.05in}  & {\hskip 0.05in} Fe I & 50 & 0.17 & 0.23 & 0.29 & \nodata & \nodata \\ 
Fe II & 2 & 0.23 & 0.16 & 0.28 & 0.02 & 0.23 {\hskip 0.05in}  & {\hskip 0.05in} Fe II & 7 & 0.17 & 0.26 & 0.31 & 0.03 & 0.17 \\ 
Sr II & 1 & \nodata & \nodata & \nodata  & \nodata & \nodata {\hskip 0.05in}  & {\hskip 0.05in} Sr II & 1 & 0.65 & 0.31 & 0.72 & 0.21 & 0.68 \\ 
Ba II & 1 & \nodata & \nodata & \nodata & \nodata & \nodata {\hskip 0.05in}  & {\hskip 0.05in} Ba II & \nodata & \nodata & \nodata & \nodata & \nodata \\ 
Eu I & 1 & \nodata & \nodata & \nodata & \nodata & \nodata {\hskip 0.05in}  & {\hskip 0.05in} Eu I & \nodata & \nodata & \nodata & \nodata & \nodata \\ 
\\
\multicolumn{7}{c}{\textbf{TucII-305}} &
\multicolumn{7}{c}{\textbf{TucII-306}}\\
CH & 2 & 0.26 & 0.44 & 0.51 & 0.25 & 0.36 {\hskip 0.05in}  & {\hskip 0.05in} CH & 2 & 0.22 & 0.51 & 0.56 & 0.27 & 0.35 \\ 
Na I & 2 & 0.26 & 0.18 & 0.32 & 0.05 & 0.26 {\hskip 0.05in}  & {\hskip 0.05in} Na I & 2 & 0.22 & 0.25 & 0.33 & 0.06 & 0.23 \\ 
Mg I & 3 & 0.26 & 0.17 & 0.31 & 0.09 & 0.28 {\hskip 0.05in}  & {\hskip 0.05in} Mg I & 6 & 0.22 & 0.30 & 0.37 & 0.21 & 0.30 \\ 
Al I & 2 & \nodata & \nodata & \nodata & \nodata & \nodata {\hskip 0.05in}  & {\hskip 0.05in} Al I & 2 & \nodata & \nodata & \nodata & \nodata & \nodata \\ 
Si I & 1 & \nodata & \nodata & \nodata & \nodata & \nodata {\hskip 0.05in}  & {\hskip 0.05in} Si I & 2 & \nodata & \nodata & \nodata & \nodata & \nodata \\ 
Ca I & 2 & 0.42 & 0.19 & 0.46 & 0.04 & 0.42 {\hskip 0.05in}  & {\hskip 0.05in} Ca I & 4 & 0.22 & 0.20 & 0.30 & 0.09 & 0.24 \\ 
Sc II & 4 & 0.26 & 0.12 & 0.29 & 0.10 & 0.28 {\hskip 0.05in}  & {\hskip 0.05in} Sc II & 5 & 0.22 & 0.22 & 0.31 & 0.12 & 0.25 \\ 
Ti II & 11 & 0.26 & 0.11 & 0.28 & 0.09 & 0.28 {\hskip 0.05in}  & {\hskip 0.05in} Ti II & 15 & 0.27 & 0.18 & 0.32 & 0.08 & 0.28 \\ 
Cr I & 2 & 0.26 & 0.22 & 0.34 & 0.03 & 0.26 {\hskip 0.05in}  & {\hskip 0.05in} Cr I & 3 & 0.22 & 0.30 & 0.37 & 0.03 & 0.22 \\ 
Mn I & 3 & 0.26 & 0.35 & 0.44 & 0.20 & 0.33 {\hskip 0.05in}  & {\hskip 0.05in} Mn I & 2 & 0.22 & 0.79 & 0.82 & 0.65 & 0.69 \\ 
Fe I & 41 & 0.26 & 0.23 & 0.35 & \nodata & \nodata {\hskip 0.05in}  & {\hskip 0.05in} Fe I & 59 & 0.22 & 0.28 & 0.36 & \nodata & \nodata \\ 
Fe II & 2 & 0.26 & 0.09 & 0.28 & 0.05 & 0.26 {\hskip 0.05in}  & {\hskip 0.05in} Fe II & 4 & 0.22 & 0.14 & 0.26 & 0.03 & 0.22 \\ 
Ni I & 1 & 0.26 & 0.20 & 0.33 & 0.06 & 0.27 {\hskip 0.05in}  & {\hskip 0.05in} Co I & 2 & 0.42 & 0.30 & 0.52 & 0.06 & 0.42 \\ 
Sr II & 1 & \nodata & \nodata & \nodata & \nodata & \nodata {\hskip 0.05in}  & {\hskip 0.05in} Sr II & 1 & 0.30 & 0.22 & 0.37 & 0.12 & 0.32 \\ 
Ba II & 1 & \nodata & \nodata & \nodata & \nodata & \nodata {\hskip 0.05in}  & {\hskip 0.05in} Ba II & 1 & \nodata & \nodata & \nodata & \nodata & \nodata \\ 
Eu I & 1 & \nodata & \nodata & \nodata & \nodata & \nodata {\hskip 0.05in}  & {\hskip 0.05in} Eu I & 1 & \nodata & \nodata & \nodata & \nodata & \nodata \\ 
\\
\multicolumn{7}{c}{\textbf{TucII-309}}\\
CH & 2 & 0.22 & 0.24 & 0.33 & 0.20 & 0.30 {\hskip 0.05in}\\ 
Na I & 2 & 0.22 & 0.27 & 0.35 & 0.11 & 0.25 {\hskip 0.05in}\\ 
Mg I & 5 & 0.22 & 0.23 & 0.32 & 0.11 & 0.25 {\hskip 0.05in}\\ 
Al I & 1 & \nodata & \nodata & \nodata & \nodata & \nodata {\hskip 0.05in}\\ 
Si I & 2 & \nodata & \nodata & \nodata & \nodata & \nodata {\hskip 0.05in}\\ 
Ca I & 15 & 0.22 & 0.18 & 0.28 & 0.10 & 0.24 {\hskip 0.05in}\\ 
Sc II & 2 & 0.33 & 0.51 & 0.61 & 0.35 & 0.48 {\hskip 0.05in}\\ 
Ti I & 6 & 0.22 & 0.27 & 0.35 & 0.05 & 0.23 {\hskip 0.05in}\\ 
Ti II & 18 & 0.24 & 0.25 & 0.35 & 0.07 & 0.25 {\hskip 0.05in}\\ 
Cr II & 1 & 0.40 & 0.25 & 0.47 & 0.07 & 0.41 {\hskip 0.05in}\\ 
Cr I & 13 & 0.23 & 0.25 & 0.34 & 0.01 & 0.23 {\hskip 0.05in}\\ 
Mn I & 3 & 0.22 & 0.48 & 0.53 & 0.24 & 0.33 {\hskip 0.05in}\\ 
Fe I & 126 & 0.22 & 0.25 & 0.33 & \nodata & \nodata {\hskip 0.05in}\\ 
Fe II & 10 & 0.33 & 0.26 & 0.42 & 0.05 & 0.33 {\hskip 0.05in}\\ 
Sr II & 1 & 0.50 & 0.43 & 0.66 & 0.50 & 0.71 {\hskip 0.05in}\\ 
Ba II & 1 & \nodata & \nodata & \nodata & \nodata & \nodata {\hskip 0.05in}\\ 
Eu I & 1 & \nodata & \nodata & \nodata & \nodata & \nodata {\hskip 0.05in}\\ 
\enddata
\end{deluxetable*}

\startlongtable
\begin{deluxetable*}{lllrrrr} 
\tablecolumns{6}
\footnotesize
\tablewidth{\textwidth}
\tablecaption{Line measurements\label{tab:linetable}}
\tablehead{   
  \colhead{Star} &
  \colhead{Rest} &
  \colhead{Species} & 
  \colhead{Excitation} & 
  \colhead{Oscillator} &
  \colhead{Equivalent} & 
  \colhead{$\log\epsilon (\rm{X})$} \\
\colhead{name} &
\colhead{wavelength (\AA)} &
  \colhead{} & 
  \colhead{potential (eV)} & 
  \colhead{strength} &
  \colhead{width (m\AA)} & 
  \colhead{} 
}
\startdata
TucII-301 & 4313.00 & CH & syn & syn & syn & 5.49 \\
TucII-301 & 4323.00 & CH & syn & syn & syn & 5.79 \\
TucII-301 & 5889.95 & Na I & 0.00 & 0.11 & 117.5 & 3.19 \\
TucII-301 & 5895.92 & Na I & 0.00 & $-$0.19 & 78.2 & 2.73 \\
TucII-301 & 3832.30 & Mg I & 2.71 & 0.27 & 152.6 & 4.52 \\
TucII-301 & 3838.29 & Mg I & 2.72 & 0.49 & 145.8 & 4.22 \\
TucII-301 & 5172.68 & Mg I & 2.71 & $-$0.45 & 148.7 & 4.88 \\
TucII-301 & 5183.60 & Mg I & 2.72 & $-$0.24 & 195.7 & 5.18 \\
\enddata
\tablecomments{Table~\ref{tab:linetable} is published in its entirety in the machine-readable format.
      A portion is shown here for guidance regarding its form and content.}
\end{deluxetable*}

\appendix

\section{Compilation of velocity measurements of Tucana II members}
\label{app:vels}

In Table~\ref{tab:velocities}, we present a compilation of all velocity measurements of Tucana II members in the literature. 
A subset of these measurements were used for the dynamical analysis in this paper; see Section~\ref{sec:velocities} for more details. 
Results of the dynamical analysis are presented in Section~\ref{sec:dynamics}.

\startlongtable
\begin{deluxetable*}{lllrrrrrr} 
\tablecolumns{6}
\footnotesize
\tablewidth{\textwidth}
\tablecaption{Velocity measurements\label{tab:velocities}}
\tablehead{   
  \colhead{Name} &
  \colhead{R.A. (deg)} &
  \colhead{DEC (deg)} & 
  \colhead{MJD\tablenotemark{a}} & 
  \colhead{Instrument} &
  \colhead{$v_{\text{helio}}$} & 
  \colhead{$v_{\text{err}}$} &
  \colhead{Ref.}\\
\colhead{} &
\colhead{(J2000)} &
  \colhead{(J2000)} & 
  \colhead{} & 
  \colhead{} &
  \colhead{(km\,s$^{-1}$)} & 
  \colhead{(km\,s$^{-1}$)}  & 
}
\startdata
TucII-006 & 342.92942 & $-$58.54269 & 57221.3 & M2FS & $-126.0$\tablenotemark{b} & 0.7 & \citet{wmo+16} \\
\nodata & 342.92942 & $-$58.54269 & 57278.0 & M2FS & $-125.4$\tablenotemark{b} & 0.8 & \citet{wmo+16} \\
\nodata & 342.92942 & $-$58.54269 & 57228.0 & IMACS & $-126.4$ & 1.0 & \citet{cfs+21} \\
\nodata & 342.92942 & $-$58.54269 & 57534.4 & IMACS & $-125.0$ & 1.6 & \citet{cfs+21} \\
\nodata & 342.92942 & $-$58.54269 & 57638.2 & IMACS & $-127.6$ & 1.2 & \citet{cfs+21} \\
\nodata & 342.92942 & $-$58.54269 & 57629.7 & MIKE & $-123.3$ & 1.2 & Re-measured from \citet{jfe+16} \\
\nodata & 342.92942 & $-$58.54269 & 57981.2 & MIKE & $-126.4$ & 1.3 & Re-measured from \citet{cfj+18} \\
TucII-011 & 342.95950 & $-$58.62783 & 57221.3 & M2FS & $-125.1$ & 0.5  & \citet{wmo+16} \\
\nodata & 342.95950 & $-$58.62783 & 57278.0 & M2FS & $-124.5$\tablenotemark{b} & 0.5 & \citet{wmo+16} \\
\nodata & 342.95950 & $-$58.62783 & 57228.0 & IMACS & $-126.6$ & 1.0 & \citet{cfs+21} \\
\nodata & 342.95950 & $-$58.62783 & 57534.4 & IMACS & $-127.4$ & 1.5 & \citet{cfs+21} \\
\nodata & 342.95950 & $-$58.62783 & 57638.2 & IMACS & $-126.0$ & 1.2 & \citet{cfs+21} \\
\nodata & 342.95950 & $-$58.62783 & 57629.7 & MIKE & $-124.1$ & 1.2 & Re-measured from \citet{jfe+16}\\
\nodata & 342.95950 & $-$58.62783 & 57981.1 & MIKE & $-125.7$ & 1.2 & Re-measured from \citet{cfj+18}\\
TucII-022 & 343.08908 & $-$58.51869 & 57221.3 & M2FS & $-117.7$\tablenotemark{b} & 2.0 & \citet{wmo+16}\\
\nodata & 343.08908 & $-$58.51869 & 57228.0 & IMACS & $-$120.8 & 1.1 & \citet{cfs+21} \\
\nodata & 343.08908 & $-$58.51869 & 57534.4 & IMACS & $-$120.6 & 2.7 & \citet{cfs+21} \\
\nodata & 343.08908 & $-$58.51869 & 57638.2 & IMACS & $-$121.3 & 1.3 & \citet{cfs+21} \\
\nodata & 343.08908 & $-$58.51869 & 58699.4 & MagE & $-$123.3 & 3.3 & \citet{cfs+21} \\
TucII-033 & 342.78467 & $-$58.55225 & 57221.3 & M2FS & $-123.9$\tablenotemark{b} & 0.5 & \citet{wmo+16}\\
\nodata & 342.78467 & $-$58.55225 & 57278.0 & M2FS & $-126.0$\tablenotemark{b} & 0.5 & \citet{wmo+16}\\
\nodata  & 342.78467 & $-$58.55225 & 57630.2 & MIKE & $-126.4$ & 1.2 & Re-measured from \citet{jfe+16} \\
\nodata  & 342.78467 & $-$58.55225 & 57981.3 & MIKE & $-126.3$ & 1.2 & Re-measured from \citet{cfj+18} \\
\nodata  & 342.78467 & $-$58.55225 & 58701.2 & MagE & $-$129.1 & 3.1 & Re-measured from \citet{cfj+18} \\
TucII-052 & 342.71513 & $-$58.57569 & 57221.3 & M2FS & $-120.3$\tablenotemark{b} & 0.7 & \citet{wmo+16} \\
\nodata & 342.71513 & $-$58.57569 & 57278.0 & M2FS & $-121.4$\tablenotemark{b} & 0.9 & \citet{wmo+16} \\
\nodata & 342.71513 & $-$58.57569 & 57630.2 & MIKE & $-121.1$ & 1.2 & Re-measured from \citet{jfe+16} \\
\nodata & 342.71513 & $-$58.57569 & 57981.2 & MIKE & $-121.1$ & 1.2 & Re-measured from \citet{cfj+18} \\
TucII-074 & 343.27779 & $-$58.52111 & 57221.3 & M2FS & $-123.5$\tablenotemark{b} & 1.5 & \citet{wmo+16} \\
\nodata & 343.27779 & $-$58.52111 & 57278.0 & M2FS & $-127.8$\tablenotemark{b} & 1.8 & \citet{wmo+16} \\
TucII-078 & 342.67112 & $-$58.51897 & 57221.3 & M2FS & $-132.5$\tablenotemark{b} & 0.9 & \citet{wmo+16} \\
\nodata & 342.67112 & $-$58.51897 & 57278.0 & M2FS & $-133.3$\tablenotemark{b} & 0.8 & \citet{wmo+16} \\
\nodata & 342.67112 & $-$58.51897 & 57980.1 & MIKE & $-124.5$ & 1.2 & Re-measured from \citet{cfj+18} \\
TucII-085 & 343.31625 & $-$58.53128 & 57221.3 & M2FS & $-128.3$\tablenotemark{b} & 8.9 & \citet{wmo+16} \\
Star12 & 342.87278 & $-$58.51841 & 57228.0 & IMACS & $-$128.3 & 2.8 & \citet{cfs+21} \\
\nodata & 342.87278 & $-$58.51841 & 57534.4 & IMACS & $-$135.1 & 5.2 & \citet{cfs+21} \\
\nodata & 342.87278 & $-$58.51841 & 57638.2 & IMACS & $-$134.8 & 1.8 & \citet{cfs+21} \\
Star68 & 343.13634 & $-$58.60846 & 57228.0 & IMACS & $-$128.0 & 1.2 & \citet{cfs+21} \\
\nodata & 343.13634 & $-$58.60846 & 57534.4 & IMACS & $-$131.6 & 2.2 & \citet{cfs+21} \\
\nodata & 343.13634 & $-$58.60846 & 57638.2 & IMACS & $-$126.5 & 1.4 & \citet{cfs+21} \\
\nodata & 343.13634 & $-$58.60846 & 58701.2 & MagE & $-$123.0 & 3.3 & \citet{cfs+21} \\
TucII-203 & 342.53696 & $-$58.49975 & 57979.6 & MIKE & $-126.3$ & 1.2 & Re-measured from \citet{cfj+18} \\
TucII-206 & 343.65279 & $-$58.61608 & 58036.5 & MIKE & $-122.8$ & 1.2 & Re-measured from \citet{cfj+18} \\
TucII-301 & 342.68790 & $-$58.93902 & 58701.0 & MagE & $-$128.0 & 3.3 & \citet{cfs+21} \\
\nodata & 342.68790 & $-$58.93902 & 59426.3 & MIKE & $-125.5$ & 0.9 & This work \\
\nodata & 342.68790 & $-$58.93902 & 59493.2 & MIKE & $-122.6$ & 1.1 & This work \\
TucII-303 & 343.27164 & $-$57.90751 & 58700.1 & MagE & $-$130.0 & 3.5 & \citet{cfs+21} \\
\nodata & 343.27164 & $-$57.90751 & 59133.3 & MIKE & $-128.7$ & 0.9 & This work \\
\nodata & 343.27164 & $-$57.90751 & 59185.1 & MIKE & $-129.4$ & 0.9 & This work \\
TucII-305 & 344.44525 & $-$57.72758 & 58701.3 & MagE & $-$124.5 & 3.1 & \citet{cfs+21} \\
\nodata & 344.44525 & $-$57.72758 & 59132.2 & MIKE & $-125.3$ & 0.9 & This work \\
TucII-306 & 342.90425 & $-$58.89377 & 58700.1 & MagE & $-$120.2 & 3.1 & \citet{cfs+21} \\
\nodata & 342.90425 & $-$58.89377 & 59426.4 & MIKE & $-118.8$ & 0.9 & This work \\
\nodata & 342.90425 & $-$58.89377 & 59493.1 & MIKE & $-119.2$ & 0.9 & This work \\
TucII-309 & 342.35287 & $-$58.34651 & 58700.3 & MagE & $-$133.8 & 3.1 & \citet{cfs+21} \\
\nodata & 342.35287 & $-$58.34651 & 59133.1 & MIKE & $-126.1$ & 0.9 & This work \\
\nodata & 342.35287 & $-$58.34651 & 59373.2 & MIKE & $-124.0$ & 0.9 & This work \\
\nodata & 342.35287 & $-$58.34651 & 59493.2 & MIKE & $-125.0$ & 0.9 & This work \\
TucII-310 & 343.19740 & $-$58.76781 & 58700.2 & MagE & $-$124.6 & 3.5 & \citet{cfs+21}\\
TucII-320 & 342.75384 & $-$58.53725 & 58699.3 & MagE & $-$115.6 & 3.2 & \citet{cfs+21}\\
\enddata
\tablenotetext{a}{Defined as the MJD at the midpoint of observation. For velocities reported in \citet{wmo+16}, we list the MJD derived from in Table 1 in that study.}
\tablenotetext{b}{Offset of +2.5\,km\,s$^{-1}$ has been applied to account for a zero-point offset between M2FS and MIKE velocities (see paragraph 4 in Section~\ref{sec:velocities}).}
\vspace{-2em}
\end{deluxetable*}

\end{document}